\documentclass[%
aps,reprint,preprintnumbers,nofootinbib
]{revtex4-2}

\usepackage[colorlinks = true, linkcolor = blue, urlcolor  = blue, citecolor = blue, anchorcolor = blue]{hyperref}
\usepackage[T1]{fontenc}
\usepackage[latin9]{inputenc}
\usepackage{babel}
\usepackage{nccmath}
\usepackage{orcidlink}
\usepackage{placeins}
\usepackage{subcaption}
\usepackage[usenames,dvipsnames]{xcolor}

\usepackage[bb=boondox]{mathalfa}  

\usepackage{graphicx,amsmath,amssymb,color}
\usepackage[normalem]{ulem}
\usepackage{amsfonts}
\usepackage[toc,page]{appendix}
\usepackage{url} 
\usepackage{latexsym}
\usepackage{algpseudocode}
\usepackage{amsthm}
\usepackage{mathrsfs}
\usepackage{natbib}
\usepackage{color,verbatim}
\usepackage{psfrag}
\usepackage{relsize}
\usepackage{tikz}
\usepackage{etoolbox}
\newcommand{\circled}[2][]{%
  \tikz[baseline=(char.base)]{%
    \node[shape = circle, draw, inner sep = 1pt]
    (char) {\phantom{\ifblank{\smaller \smaller #1}{\smaller \smaller #2}{\smaller \smaller #1}}};%
    \node at (char.center) {\makebox[0pt][c]{\smaller \smaller #2}};}}
\robustify{\circled}

\newcommand{\vev}[1]{\langle #1\rangle}

\newcommand{\ket}[1]{|#1 \rangle}
\newcommand{\braket}[2]{\left \langle #1 \mid #2 \right \rangle}

\def\beq{\begin{equation}}
\def\eeq{\end{equation}}
\def\bali{{\begin{align}}}
\def\eali{{\end{align}}}
\def\ie{{\it i.e.~}}

\newcommand{\fv}{\varphi_{\rm fv}}
\newcommand{\tv}{\varphi_{\rm tv}}

\newcommand{\cO}{{\cal O}}
\newcommand{\cU}{{\cal U}}
\newcommand{\scV}{{\mathscr{V}}}

\usepackage[labelfont=bf,labelsep=colon]{caption}
\gdef\@fpheader{}

\begin{document}

\preprint{IPPP/25/37} 

\title{Qumode Tensor Networks for False Vacuum Decay in Quantum Field Theory}

\author{
Steven Abel$^{(1,2)}$\orcidlink{0000-0003-1213-907X}}
\email{s.a.abel@durham.ac.uk}

\author{Michael Spannowsky$^{(1)}$\,\orcidlink{0000-0002-8362-0576}}
\email{michael.spannowsky@durham.ac.uk}

\author{Simon Williams$^{(1)}$\,\orcidlink{0000-0001-8540-0780}}
\email{simon.j.williams@durham.ac.uk}

\address{\vspace{0.5cm} 
\hspace{-0.2cm}$^{(1)}$Institute for Particle Physics Phenomenology, \\Durham University, Durham DH1 3LE, UK}
\address{\vspace{0.2cm}$^{(2)}$Department of Mathematical Sciences, \\Durham University, Durham DH1 3LE, UK}

\begin{abstract}\vspace{0.2cm}False vacuum decay in scalar quantum field theory (QFT) is a cornerstone of early Universe cosmology and high energy  physics, yet its real-time dynamics is essentially inaccessible to classical computation due to its non-perturbative, highly entangled dynamics. We introduce a general Hamiltonian framework for simulating full interacting QFTs, using a spatial lattice of continuos-variable ``qumodes'' -- bosonic local oscillators whose high-dimensional local Hilbert space faithfully captures interacting field dynamics. This construction is rooted in continuous-variable quantum computing (CVQC), and provides a unified platform spanning efficient classical tensor-network methods and emerging photonic quantum hardware.
The first key advance of this work is a robust and scaleable method for preparing the QFT in its correct initial vacuum state. We develop an imaginary-time preparation algorithm tailored to  qumode lattices, that efficiently projects onto the  vacuum even in strongly coupled regimes. This provides  a controllable starting point for studying nonperturbative dynamics such as tunnelling and real-time decay.
Building on this, we use a time-evolving block decimation algorithm to capture the real-time dynamics of the scalar field. 
Our second key advance is the identification and excitation of the negative fluctuation mode of the bounce configuration on the qumode lattice. A small displacement along this mode produces the expected tachyonic growth, driving fully coherent bubble nucleation without requiring classically supercritical seeds. This demonstrates that the qumode lattice captures non-perturbative quantum dynamics that lie beyond the classical treatments.
Our results  establish the qumode network as a scalable framework for non-equilibrium scalar QFT phenomena and pave the way for higher-dimensional studies and continuous-variable quantum computing implementations.
\end{abstract}

\maketitle


\begin{figure*}[t]
    \centering
    \includegraphics[width=\textwidth]{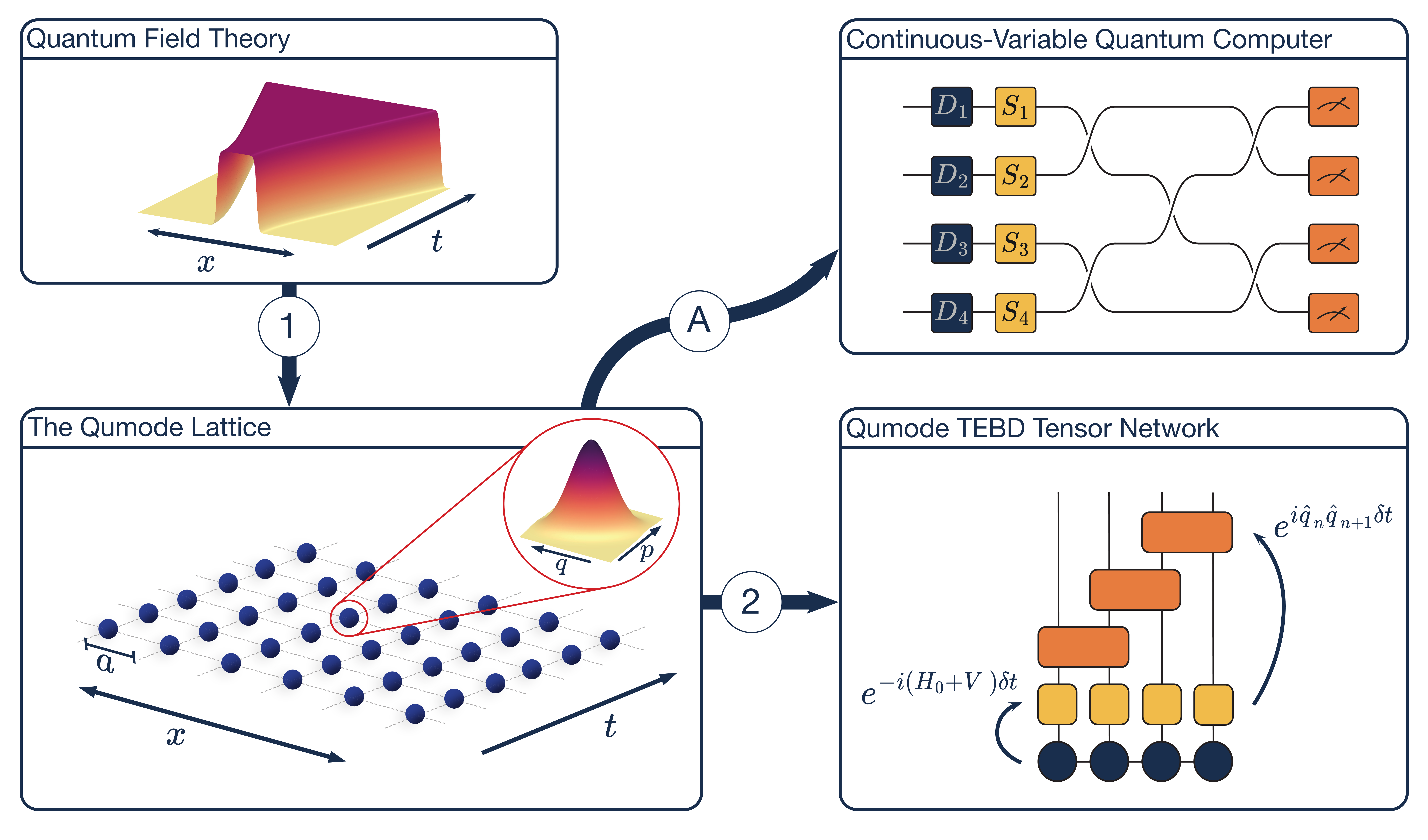}
    \caption{Schematic overview of simulating quantum field theories with the qumode lattice. A (1+1)-dimensional quantum field theory, with field dynamics governed by a local potential $V(\phi)$, is \raisebox{0.15ex}{\circled[A]{1}} discretised in space to form a lattice of qumodes, where each site encodes the field and its conjugate momentum as continuous quadratures $\hat{q}$ and $\hat{p}$, respectively. This qumode lattice can be simulated in two ways: \raisebox{0.15ex}{\circled[A]{A}} directly on a continuous-variable quantum computer, as outlined in Ref.~\cite{Abel:2025zxb}; or \raisebox{0.15ex}{\circled[A]{2}} emulated classically using a qumode TEBD tensor network, where time evolution is approximated by alternating layers of single- and two-mode gates.}
    \label{fig:fullProcess}
\end{figure*}

\section{Introduction}

The purpose of this paper is to establish the first framework that enables direct, real-time simulation of non-perturbative processes, within a full scalar quantum field theory (QFT). By formulating the problem in the Hamiltonian picture and representing the field as a lattice of continuous-variable modes, our approach captures non-perturbative dynamics beyond the reach of semiclassical and path-integral techniques. Furthermore, by exploiting continuous-variable quantum computing (CVQC) techniques, this approach provides an efficient route towards quantum simulation of scalar QFTs, reducing the required quantum volume in comparison to discrete-variable quantum computing (DVQC) platforms. In this paper, we focus on simulating the real-time dynamics of false-vacuum decay, establishing a new route to studying phase transitions from first principles, bridging fundamental quantum field dynamics and quantum simulation platforms.

False-vacuum decay in scalar QFT is a cornerstone in our understanding of the dynamics of phase transitions. These processes have profound implications for cosmology, particle physics, and the evolution of the early universe. Indeed, it is thought that they  played a central role in setting the initial conditions of the universe, influencing both the large-scale structure of the cosmos and the distribution of matter. Moreover, at least one such transition, namely the electroweak phase transition, is {\it known} to have taken place in the early universe. This event could provide explanations for the observed matter-antimatter asymmetry and could generate observable gravitational wave signatures.  Thus, these mechanisms provide a crucial theoretical bridge linking microphysical interactions and cosmological scales, 
and understanding them is essential for reconstructing the history of the cosmos.

At its core, a phase transition in scalar QFT involves the probabilistic escape of the quantum field from a metastable state, the false vacuum, to a lower energy configuration, the true vacuum, by tunnelling through an energy barrier in the potential.
This leads to the nucleation of spherical bubbles of true vacuum, which then grow to fill the entire system as the false vacuum ``boils off''.
The study of this process was pioneered in the seminal works of Callan, Coleman and de Luccia~\cite{PhysRevD.15.2929,PhysRevD.16.1762,1980PhRvD..21.3305C}, who provided a robust semi-analytical framework for the process.
Their analysis focussed on what happens during the nucleation of a single bubble, using the spherical symmetry to reduce the vacuum decay to a one-dimensional quantum mechanical problem, in which the transition pathway between vacua is modelled through the so-called {\it bounce solution}.  From the perspective of four-dimensional space-time this solution corresponds to the instantaneous production of an extended  ``lump'' of the quantum field known as an {\it instanton}, which essentially seeds the nucleating bubble of true vacuum.

This seminal work provided a conceptual framework for analysing the rate of bubble nucleation and it forms the foundation of virtually all the subsequent work that invokes phase transitions in scalar QFT. However, the picture it paints also tells us why in the intervening decades it has never been possible to {\it simulate}  phase transitions in scalar QFT. The source of the difficulty is the following fact: despite tunnelling being a quantum phenomenon, the large instanton configuration which drives the whole process can somewhat counterintuitively be thought of as a classical object.
It is ``classical'' in the sense that the bounce solution is the exponentially dominant contribution to the tunnelling process, which merely gets dressed by quantum fluctuations.
However, this classical lump straddles the false and true vacua and does not belong in either. Consequently, from the perspective of the system sitting in the false vacuum, the instanton configuration that it needs to reach in order to tunnel represents a large coherent state with a huge degree of entanglement. Indeed, thanks to the symmetry of instanton solutions, we can state quite categorically that the longer the lifetime of the false vacuum, the greater the degree of entanglement required to describe its decay. This statement can be formulated more precisely: the entanglement entropy of an instanton configuration is expected to obey an area\,$\Leftrightarrow$\,entropy law, $S_{\text{ent}} \sim \frac{A}{a^{d-2}}
$, where $A$ is the surface area of the solution and $a$ is some kind of short-distance cut-off. But the action of the bounce-solution also depends on $A$ as $S \sim \sigma \times A$ where $\sigma $ is the surface tension, such that the rate of bubble formation is proportional to $\Gamma \propto e^{-\sigma A}$.

In summary, then, quantum entanglement is integral to tunnelling in scalar QFT because it underpins the coherence of quantum states across spatial regions. This coherence is essential for the collective behaviour that characterises the formation of nucleating bubbles, and yet it also makes simulating tunnelling in scalar QFT very challenging: one must effectively remain in the false vacuum while generating sufficient entanglement to produce a coherent instanton state. There have been several studies which have sought to simulate aspects of the tunnelling process, which have provided valuable insights~\cite{Abel:2020qzm, Abel:2020ebj, Ng:2020pxk, PRXQuantum.3.020316, Szasz-Schagrin:2022wkk, Lagnese:2021grb, Pirvu:2023plk, Zenesini:2023afv, Darbha:2024srr, Vodeb:2024tvo, Luo:2025qlg}.
However, in large part due to this difficulty, they considered restricted setups, such as Ising spin-chain models, rather than a full QFT framework.

In this paper, we present an approach that, for the first time, enables the simulation of false vacuum decay within a full scalar QFT framework. Our approach utilises the Hamiltonian formulation of QFT, which bypasses the notorious ``sign problem'' that is ubiquitous in Monte Carlo techniques for solving the path integral~\cite{PhysRevLett.46.77, VONDERLINDEN199253, PhysRevE.49.3855}. One key advantage of the Hamiltonian perspective is its natural compatibility with real-time evolution, enabling the study of inherently dynamical processes~\cite{doi:10.1126/science.1217069, 10.5555/2685155.2685163, Van_Damme_2021, PRXQuantum.3.020316, Papaefstathiou:2024zsu, Bennewitz:2024ixi, Rigobello:2021fxw, Abel:2024kuv, PhysRevResearch.6.033057, Jha:2024jan, Carena:2024peb, Ale:2024uxf, zemlevskiy2024, crane2024,PhysRevResearch.6.043065, Chai2025fermionicwavepacket, Abel:2025zxb, Ingoldby:2025bdb, Davoudi:2025rdv, Schuhmacher:2025ehh}.

The Hamiltonian formulation has already proven fruitful in studying a wide array of non-equilibrium phenomena, such as quantum quenches~\cite{Martinez:2016yna, PhysRevLett.109.175302, Ingoldby:2024fcy}, thermalisation~\cite{doi:10.1126/science.abl6277, PhysRevD.106.054508, PhysRevA.108.022612, Fromm:2023npm}, and dynamic phase transitions in gauge theory~\cite{Jensen:2022hyu, Angelides:2023noe, PRXQuantum.4.030323}. Moreover, as emphasised in foundational work on Hamiltonian quantum simulation~\cite{Preskill:2018jim, Preskill:2018fag}, these methods offer a scalable route to accessing regimes far beyond current Euclidean techniques~\cite{Banuls:2019bmf, PRXQuantum.4.027001, PRXQuantum.5.037001}.

The crucial advance that we will make here is to build a {\it scalar} QFT out of a lattice of so-called {\it qumodes}, and then leverage the Hamiltonian formalism to simulate its evolution. A qumode is a large-dimensional Hilbert space that describes the quantum mechanics at a local site in space, which crucially must admit arbitrary potentials (Equivalently, it can be thought of as an arbitrary zero-dimensional field theory). By coupling such qumodes together into a lattice, we can build an entire scalar QFT that has an arbitrary potential of our choosing. This allows us to set-up and study the systems discussed above in which the scalar QFT is in a long-lived metastable state that decays to the true vacuum via bubble nucleation. 

The procedure is schematically outlined in Fig.~\ref{fig:fullProcess}. We start with the (1+1)-dimensional scalar QFT in the continuum limit, where the dynamics of the quantum field $\phi (x)$ are governed by a Hamiltonian with a free-theory contribution that is quadratic in $\phi(x)$, and a potential, $V(\phi)$. To facilitate the real-time evolution of the field, we \raisebox{0.15ex}{\circled[A]{1}} discretise space and place the QFT on a spatial lattice of $N$-sites. The QFT is consequently represented as a system of coupled qumodes, with each qumode associated to a lattice site at position $x_n = na$, where $a$ is the lattice spacing. In this qumode-lattice formulation, spatial derivatives are replaced by finite differences, resulting in nearest-neighbour couplings between qumodes, and the emergence of an effective local-potential acting on individual sites~\cite{Abel:2024kuv, Abel:2025zxb}. 

The ideal system to realise such a configuration would be \raisebox{0.15ex}{\circled[A]{A}} a continuous-variable quantum computing device (CVQC), in particular a photonic platform~\cite{RevModPhys.79.135, 10.1063/1.5115814, Bromley_2020}. CVQC devices offer natural access to bosonic degrees of freedom, long coherence times, and room-temperature operation~\cite{PhysRevLett.82.1784, Adesso2014, RevModPhys.77.513}, making them especially well suited for simulating scalar field dynamics. In principle on such devices the dimensionality of the Hilbert space of a single qumode would be infinite. Thus these platforms would provide a promising route toward scalable simulations of vacuum decay and related non-perturbative quantum processes in scalar QFT. 

Unfortunately, CVQC devices are not yet large enough to construct a lattice of any great size. Therefore in this paper we demonstrate the method using \raisebox{0.15ex}{\circled[A]{2}} a qumode tensor network, by identifying a further possible approximation, which can be achieved by truncating the Hilbert space of each of the qumodes to a finite dimension, $d$, such that qumode operations now become $(d\times d)$ matrices. Therefore, to simulate the real-time dynamics of the QFT, we construct a Time-Evolving Block Decimation (TEBD) algorithm~\cite{PhysRevLett.91.147902, PhysRevLett.93.040502}. The TEBD algorithm represents the many-body wavefunction as a Matrix Product State (MPS), which in effect compresses the expoentially large Hilbert-space amplitudes into a sequence of tensors whose bond dimensions reflect the amount of quantum entanglement. The time evolution of the nearest-neightbour Hamiltonian is then performed using the Suzuki-Trotter decomposition~\cite{pjm/1103039709, 10.1063/1.526596}, where the full time-evolution operator,
\begin{equation}
    U(t) = e^{-i H t}~,
\end{equation}
is approximated with 
\begin{equation}
    \mathcal{U} (t) = \left[ \prod_i e^ {- i H_i \delta t} \right]^{t/\delta t}~,
\end{equation}
where the Hamiltonian has been split up into a sum of non-commuting parts, $H = \sum_i H_i$. The total evolution time has been \textit{Trotterised} by discretising $t$ into small steps of $\delta t$.

The TEBD algorithm implements the Trotterised time-evolution as a series of single-site gate operations, corresponding to the diagonal, on-site contributions from the Hamiltonian, and two-site gate operations, corresponding to the nearest-neighbour interaction contribution to the Hamiltonian. After each \textit{Trotter-step} of time $\delta t$, the MPS is truncated to a manageable bond dimension. This process is shown for a single Trotter step in routine \raisebox{0.15ex}{\circled[A]{2}} of Fig.~\ref{fig:fullProcess}.

A crucial advance we make here is an algorithm for initial state preparation of the full QFT. This is done by running the same TEBD through imaginary time, which projects onto the lowest energy state, arranged to be the metastable vacuum state.

\section{Qumode Lattice QFT Framework}

\begin{figure}[t!]
\centering
\includegraphics[keepaspectratio, width=0.45\textwidth]{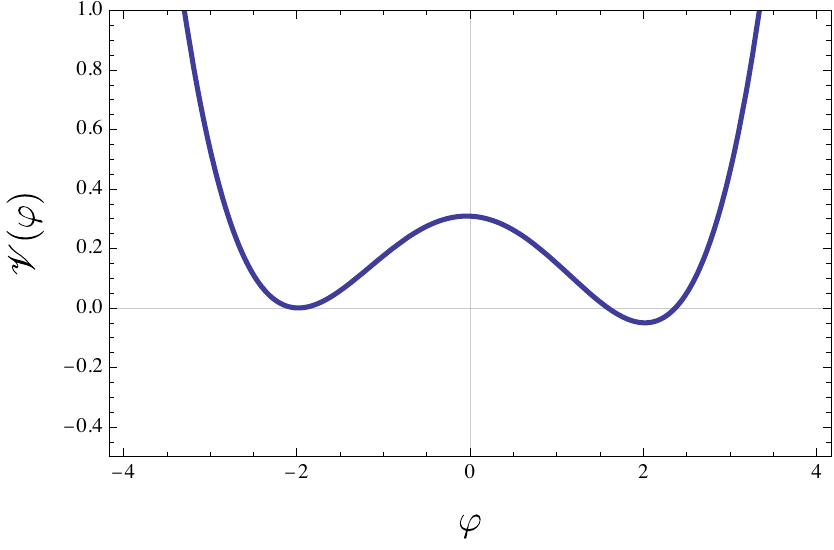}
\caption{The polynomial potential of Eq.~\eqref{eq:pot_poly}  with  $\varphi_0=2$, $\lambda=0.5$, $\varepsilon=0.1$ and $\scV_0=0.2$.}
\label{fig:pot_poly}
\end{figure}

As outlined in the Introduction we shall in this paper be interested in constructing scalar QFTs with potentials that make them unstable to decay. A proto-typical class of such theories are renormalizable scalar field theories with  potentials of the following form: 
\begin{equation}
\scV(\varphi ) ~=~ \frac{\lambda}{4!} (\varphi^2 - \varphi_0^2)^2 - \varepsilon \varphi ~-~ \scV_0 ~, 
\label{eq:pot_poly}
\end{equation}
where $\lambda$, $\varepsilon$ and $\scV_0$ are constants. 
The important feature of the potential is that it has a metastable false vacuum at $\fv \approx  -\varphi_0$. This is where we shall start the field theory. We show a concrete example with   $\lambda=0.5$, $\varepsilon=0.1$ and $\scV_0=0.2$, which gives the profile in Fig.~\ref{fig:pot_poly}. 
The field theory is expected to be able to then quantum tunnel to the true vacuum at $\tv \approx \varphi_0$  which is introduced by the second term in the potential. Note that we choose the constant $\scV_0 \approx  \varphi_0 \varepsilon $ such that the false minimum at $\fv \approx -\varphi_0 $ has zero energy,  while the true minimum at $\tv \approx \varphi_0 $ has 
\begin{equation}
\label{eq:Vp0}
    \scV(\varphi_0) ~\approx ~ - 2\varphi_0 \varepsilon ~.
\end{equation}

\subsection{Qumode preparation}

Let us begin by constructing the individual qumodes that will comprise the lattice. As described in App.~\ref{sec:basics} we use the exact correspondence between scalar free field theory and a coupled lattice of qumodes in the infinite lattice limit. 
Thus the entire field theory Hamiltonian with a lattice of $N$ qumodes is given by 
\begin{equation}
\label{eq:disc_ham}
H_{\rm QFT} ~=~ \frac{a}{2} \sum_{n=1}^N\left[   {\hat p}_n(t)^2 + 
\left(\frac{\hat q_{\,{n+1}}(t) - \hat q_n(t)}{a}\right)^2 + 2 \mathscr{V}(\hat q_n) \right]~,
\end{equation}
for arbitrary QFT potential  $\mathscr{V}(\varphi)$. Note that we do not impose periodic boundary conditions as these are not straightforward to implement on the GPU systems that we will ultimately be using: we therefore anticipate edge effects which we will have to live with. The local qumode Hamiltonian at space-position $n$ is  given by extracting the diagonal parts of the total Hamiltonian: 
\begin{equation}
\label{eq:qumode_ham}
H_{n} ~=~ \frac{a}{2} {\hat p}_n^2 + 
a^{-1} {\hat q}_{n}^2 + a \mathscr{V}(\hat q_n) ~.
\end{equation}
 We henceforth take lattice spacing $a=1$.

It is useful to study an individual qumode of the system  entirely independently as a quantum mechanical system in its own right before we tie the qumodes together into a lattice. This is a useful check of, for example, the time evolution procedure that we will be using.
To do this we will consider the diagonal Hamiltonian in Eq.~\eqref{eq:qumode_ham} but with the diagonal pieces from the $\nabla\varphi ^2 $ term subtracted: that is we will consider 
\begin{equation}
    \widetilde H_n~=~ \frac{1}{2} {\hat p}_n^2 + \mathscr{V}(\hat q_n) ~.
\end{equation}
Studying an individual qumode with this Hamiltonian will allow us to anticipate the behaviour of the full QFT, and to ensure that we are taking a sufficiently large dimensionality of the local Hilbert space. Indeed, the evolution of this individual qumode should essentially correspond to the behaviour of the entire QFT when it is oscillating in a space-independent coherent manner (as the gradient term is zero). Thus we can, for example, determine the {\it maximum} lifetime of the QFT in the false vacuum, which would correspond to the time for this individual qumode to tunnel.

In order to construct a qumode we will discretise its local Hilbert space into a large $d$-dimensional space $q_n^{j=1\ldots d}$, where
we consider eigenvalues $q_n^j \in [-L,L]$ with 
\begin{equation}
    q_n^j ~=~ -L + j\times  \delta q 
\end{equation}
where $\delta q = 2L/d$.
Thus the quantum mechanical wavefunction is given in the $\ket{q_n^i}$ basis in the obvious way: 
\begin{equation}
    \ket{\psi_n} ~=~ \sum_{i=1}^{d}   \ket{q^i_n} \braket{q^i_n}{\psi_n}~.
\end{equation}
The operator $\hat q_n $ acts diagonally on the wavefunction in this basis,  while the operator $\hat p^2 $ is an operator on the site $n$ given by finite difference in the discrete $\ket{q_n^i}$ basis. Specifically,  we have 
\begin{align}
\hat{q}_n~&=~
\begin{pmatrix}
q^1_n & 0 & \ldots \\
0 & q^j_n & \\
\vdots & & q^d_n  
\end{pmatrix}~\nonumber \\~~
    \hat{p}_n^2  ~&=~- \frac{\hbar^2}{\delta q^2} \begin{pmatrix}
 2      & -1 & ~ & ~~ & \ldots \\
-1 & 2 & -1 &~ & \ldots\\
~ & -1 & 2 & ~ & \ldots\\
\vdots & ~ & ~ & \ddots & -1 \\
\vdots & ~ & ~& ~~-1 & 2   
\end{pmatrix}
\end{align}
acting on the wavefunction  coefficients $\braket{q_n^j}{\psi_n} $. We will henceforth choose units such that $\hbar =1$. Of course, if one had access to genuine qumodes with infinite $d$ then these operators would be replaced with the corresponding continuous quadrature variable gates (for example the $\hat p_n^2$ operator would be invoked as part of a ``rotation'' gate as described in Ref.~\cite{Abel:2024kuv}).

The Hamiltonian evolution for this single qumode may be  simulated via the Schr\"odinger time-evolution operator,
\begin{equation}
     \widetilde{\mathcal{U}}_n(t) ~=~ e^{-i\widetilde H_nt}~,
\end{equation}
which propagates an initial state  $\ket{\psi_n(t=0)}$ to the state $\ket{\psi_n(t)} = \widetilde{\mathcal{U}}_n(t) \ket{\psi(0)}$ at time $t$. However it is  convenient to also  split the $\hat p_n$ and $\hat q_n$ parts, 
so that the evolution operator is approximated  by the following Trotter-Suzuki decomposition~\cite{pjm/1103039709, 10.1063/1.526596}: 
\begin{equation}
   \widetilde{\mathcal{U}}_n(t) ~=~ \left[ e^{-i\hat p_n^2 \delta t}
    e^{-i\mathscr{V}(\hat q_n) \delta t} \right]^{t/\delta t} + \mathcal{O}\left( \delta t^2\right)~. 
\end{equation}
This Trotterisation method approximates the time-evolution operator $\widetilde{ \mathcal{U}}_n(t)$ up to an error of $\mathcal{O}\left(\delta t^2\right)$, and thus is a good approximation for $\delta t \ll 1$. Trotterisation will be unavoidable later when we build the full QFT, thus using it now on a single qumode is a good opportunity to test the quality of the Trotter-Suzuki approximation in the regime of interest.

\begin{figure}[t!]
\centering
\includegraphics[keepaspectratio, width=0.45\textwidth]{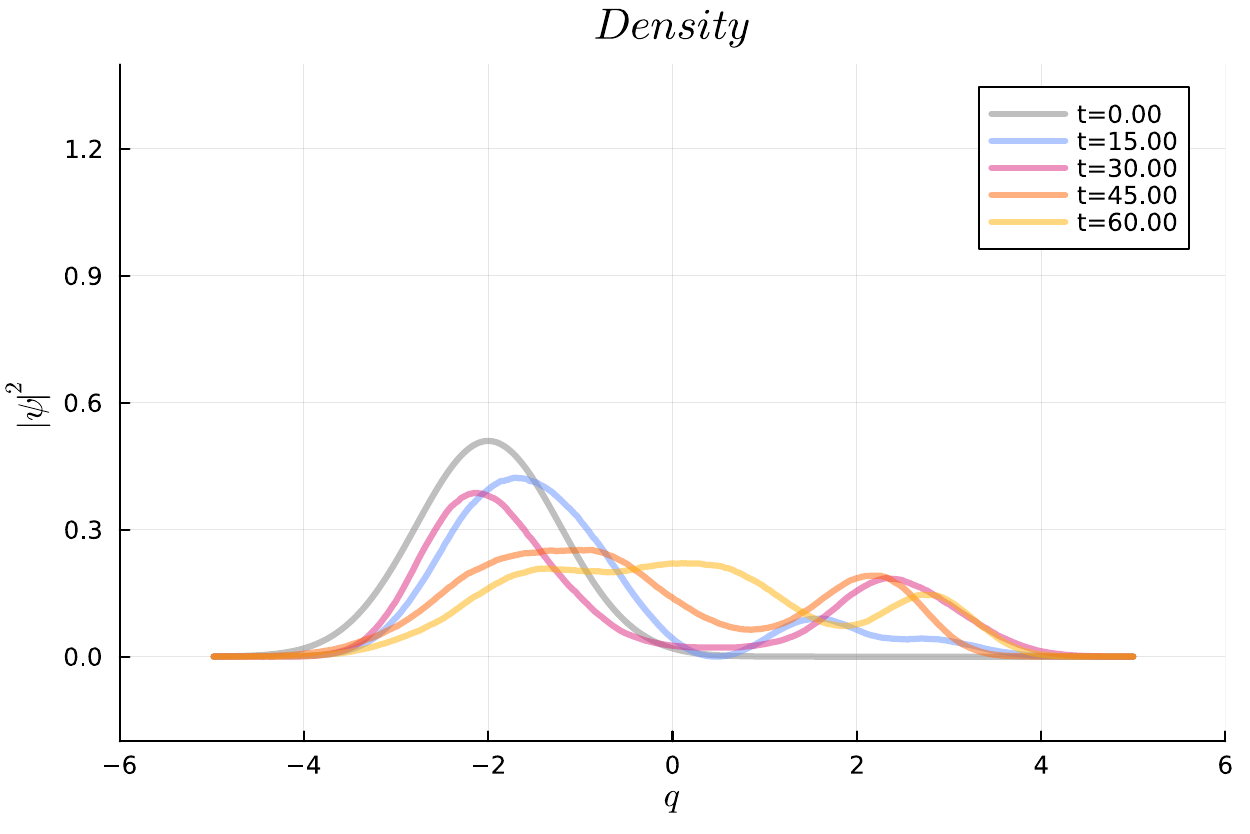}
\includegraphics[keepaspectratio, width=0.45\textwidth]{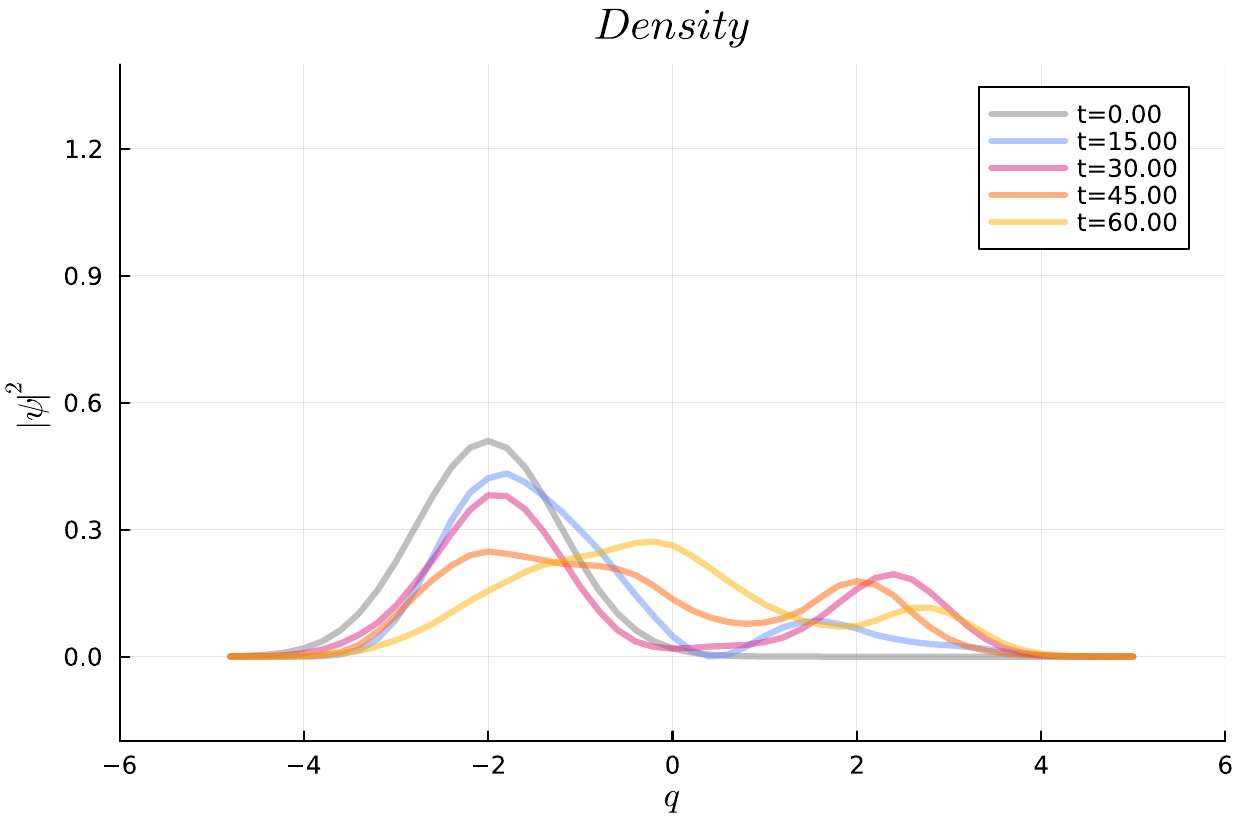}
\caption{Exact evolution of a single qumode under the potential in Fig.~\ref{fig:pot_poly} (first panel) versus a Trotterised evolution (second panel) in which the local Hilbert space is discretised as 50-dimensional, while the Trotter time-step is $\delta t=0.05$. Thus, the evolution represents up to $1200$ Trotter steps, respectively, for the five plots.}
\label{fig:testplot}
\end{figure}

In order to test the evolution we start the qumode in the ground state of the SHO with the mass corresponding to $\omega$, which is a Gaussian around the false vacuum at $\varphi = \fv = -2$. 
The evolution of a single qumode from this initial state is shown in Fig.~\ref{fig:testplot}, where in the first panel we show the evolution of the wavefunction of the continuous qumode for times up to $t=60$, simulated using \textsc{qibo} \cite{qibo_paper}, and in the second panel we show the evolution in the discretised bosonic mode with a $d=50$ dimensional Hilbert space. The Trotterised time-evolution has been performed with a time step of $\delta t= 0.05$. In the lower panel, minor deviations due to the finite truncation of the local Hilbert space are visible as a slight smoothing of the probability density at later times relative to the exact case. Despite this, it is evident that the method is resilient to Trotter error even for long evolution times. Therefore, we expect the full QFT simulations to exhibit similar robustness, with Trotter errors remaining small for sufficiently small $\delta t$.

As we will ultimately be interested in the local field expectation value, $\langle \varphi(x)  \rangle \equiv  \langle \hat q_n\rangle $, this approximation is more than adequate for our purposes. (Of course it can be improved by taking smaller Trotter steps and higher dimensionality of local Hilbert spaces.) More generally we see that the evolution displays the expected tunnelling behaviour. But what we also observe is that the tunnelling does not ``complete'' because, in what is effectively a (1+0)-dimensional field theory, the excess energy has nowhere to dissipate. As we shall see, once the qumodes are tied together into a lattice, this excess energy will ultimately be spent driving bubble expansion. 

\subsection{Qumode initialisation by adiabatic evolution}

\label{subsec:adiabat}
Now let us finesse the desired starting state for a qumode that is sitting in the false vacuum. Guided by the usual field theory discussions, one might be inclined as above to use the Simple Harmonic Oscillator (SHO) ground state of the potential expanded around the metastable minimum up to quadratic terms, with cubic and higher terms neglected. That is, if we trivially rewrite the quartic potential as an SHO potential around the false vacuum plus remaining terms as follows, 
\begin{equation}
\scV(\varphi ) ~=~ \frac{\omega^2 }{2} (\varphi -\fv)^2 + \scV_1(\varphi) 
\label{eq:pot_poly2}
\end{equation}
where in the quartic case at hand we have $\fv \approx -\varphi_0$, and hence 
\begin{equation}
    \omega^2 ~=~ \frac{\lambda \varphi_0^2}{3}~,
\end{equation}
and 
\begin{equation}
\scV_1(\varphi ) ~=~  \frac{\lambda}{4!} (\varphi + \varphi_0)^3(\varphi-3 \varphi_0) - \varepsilon \varphi ~-~ \scV_0 ~. 
\label{eq:V1}
\end{equation}
Thus, a first approximation to the desired metastable state is the ground state of the SHO with mass unity and frequency $\omega$, namely 
\begin{equation}
   \langle q_n |  \psi_n (s=0)\rangle  ~=~ \left(\frac{\omega}{\pi} \right)^{1/4} e^{-\omega  (q_n+\varphi_0)^2 /2 }~.
\end{equation}

\begin{figure}[t!]
\centering
\includegraphics[keepaspectratio, width=0.45\textwidth]{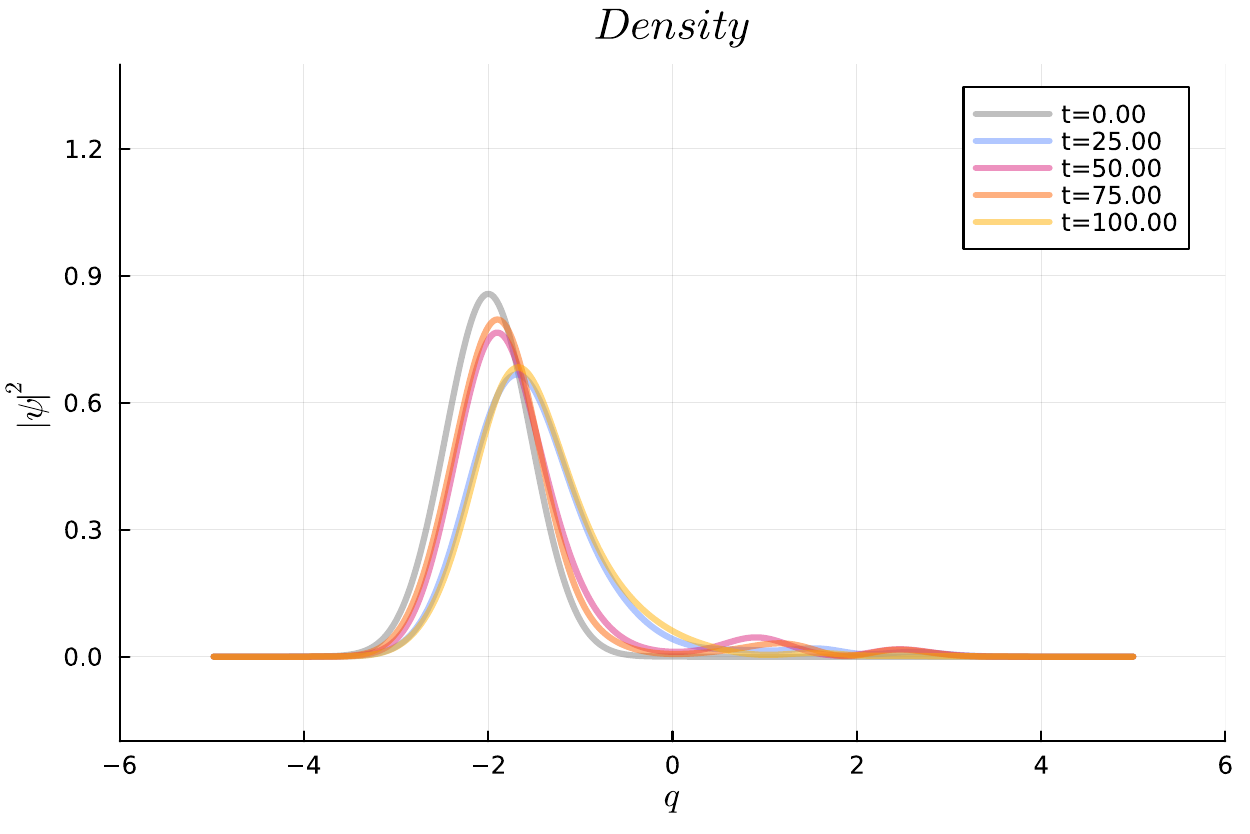}
\includegraphics[keepaspectratio, width=0.45\textwidth]{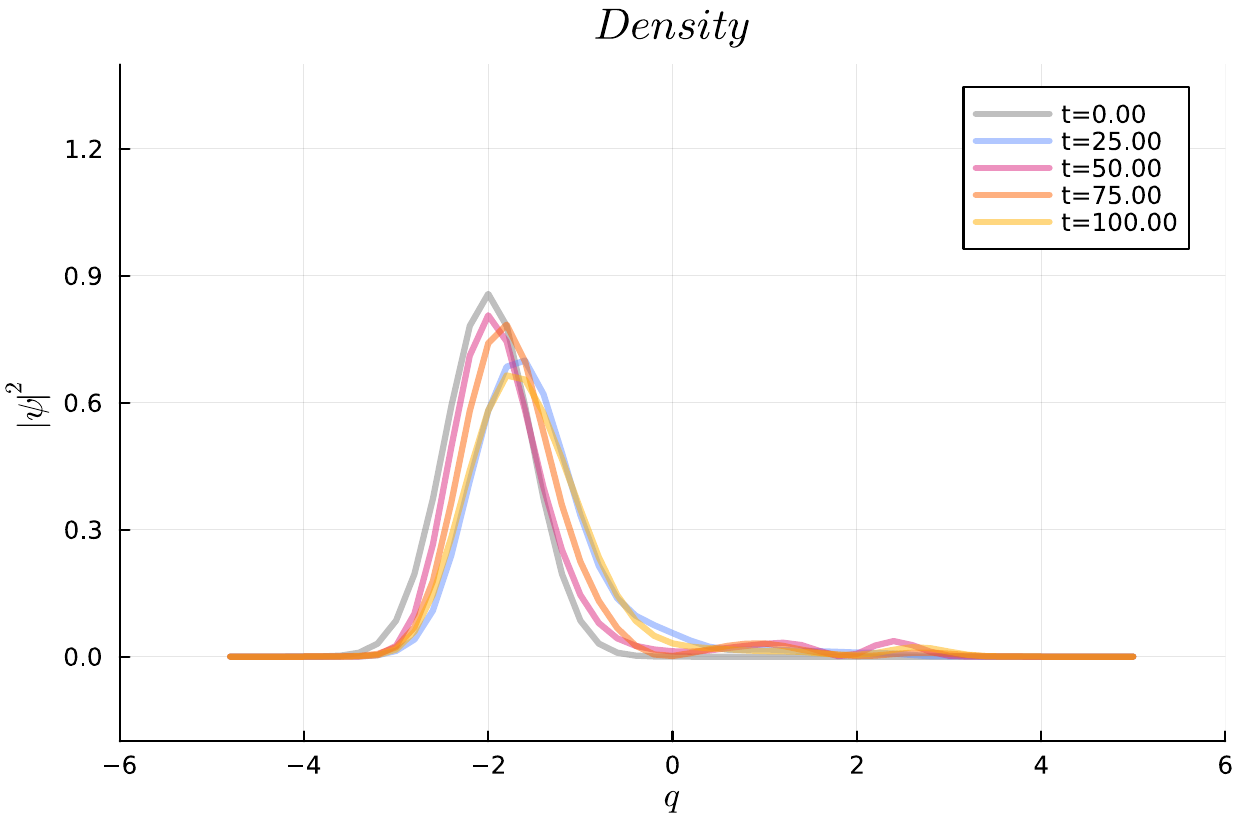}
\caption{As in Fig.~\ref{fig:testplot} but for a `stuck' single qumode with  $\lambda = 4$.}
\label{fig:testplot_stuck}
\end{figure}

Unfortunately, this approximation is a little too crude for our purposes. Indeed, inspecting the potential of Fig.~\ref{fig:pot_poly} we see that the potential well around the false vacuum at $\varphi \approx -\varphi_0$ is, unlike the SHO potential, noticeably asymmetric. As a result, a field theory that begins with all its qumodes in this SHO ground state will oscillate due to the excess energy. This can be seen in the example shown in Fig.~\ref{fig:testplot_stuck}. This qumode was initialised in the SHO ground state at $\varphi=-2$ and is ``stuck'' but continues to oscillate.
Moreover, this non-zero energy could significantly change the metastability properties when we come to construct the field theory. Furthermore one may be interested in potentials for which higher order terms dominate over the mass. For most potentials a more precise initial state for the qumodes cannot easily be determined analytically. However it can be determined using adiabatic quantum computing as suggested in ~\cite{Born_1928,farhi2000quantumcomputationadiabaticevolution, RevModPhys.90.015002, 10.1063/1.2798382, 10.1063/1.4748968}. 

\begin{figure}[t!]
\centering
\includegraphics[keepaspectratio, width=0.45\textwidth]{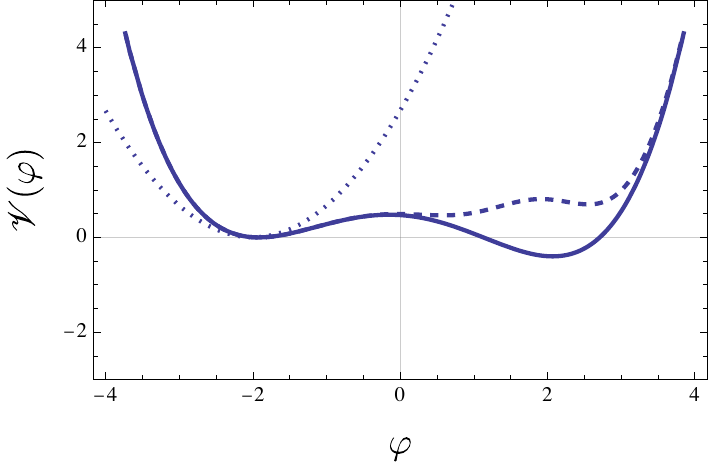}
\caption{Initialising the false vacuum state with $\lambda =1$ and $\varepsilon = 0.1$: the solid line is the full potential, the dotted line is the SHO approximation around the false vacuum where we begin the qumode in the corresponding SHO groundstate, and the dashed line is the full potential plus lift potential, where here we have taken $\delta=4$. During the adiabatic preparation of the vacuum, the potential is gradually changed from the dotted SHO potential to the dashed potential. For imaginary-time TEBD preparation of the false vacuum, the potential is set as the dashed potential. }
\label{fig:adiabatic_proc}
\end{figure}

Let us see how this can work for a single qumode. The aim is to evolve the potential from the SHO potential to an approximation to the full potential according to 
\begin{equation}
\scV_s (\hat q_n )  ~=~  \frac{\omega^2 }{2} (\hat q_n  -\fv )^2 + s \, ( \scV_1(\hat q_n ) +\scV_{\rm lift} (\hat q_n ) ) ~,
\end{equation}
where $s\in [0,1]$ is an adiabatic evolution parameter, and where $\scV_{\rm lift} (\hat q_n ) $ is a suitable potential that lifts the global minimum at $\hat q =\tv$ but does not disrupt the false minimum: for the quartic potential, we can choose  
\begin{equation}
    \scV_{\rm lift} (\hat q_n ) )~=~ ( 2 + \delta ) \, \varphi_0 \varepsilon e^{-(\hat q-\varphi_0)^2}~,
\end{equation}
which lifts the minimum at $\hat q = \varphi_0$ to $+\delta$. 
Sending the evolution parameter $s$ to one gradually reintroduces all of the remaining $\scV_1$ terms in the potential around the  false vacuum at $\hat q =\fv = -\varphi_0$, while the lift potential $\scV_{\rm lift}$ prevents the true minimum of the full potential appearing at $\hat q_n = \tv = +\varphi_0$. 
According to the adiabatic theorem, the wavefunction will remain in the ground state throughout, provided that $s$ evolves sufficiently slowly.  The lift potential ensures that the wavefunction does not simply end up in the ground state of the full potential during this process, which of course is centred around $\hat q_n =+ \varphi_0$.
A comparison of the three potentials is shown in Fig.~\ref{fig:adiabatic_proc} for the quite extreme case of $\lambda = 1$ and $\varepsilon = 0.1$.

\begin{figure}[t!]
\centering
\includegraphics[keepaspectratio, width=0.45\textwidth]{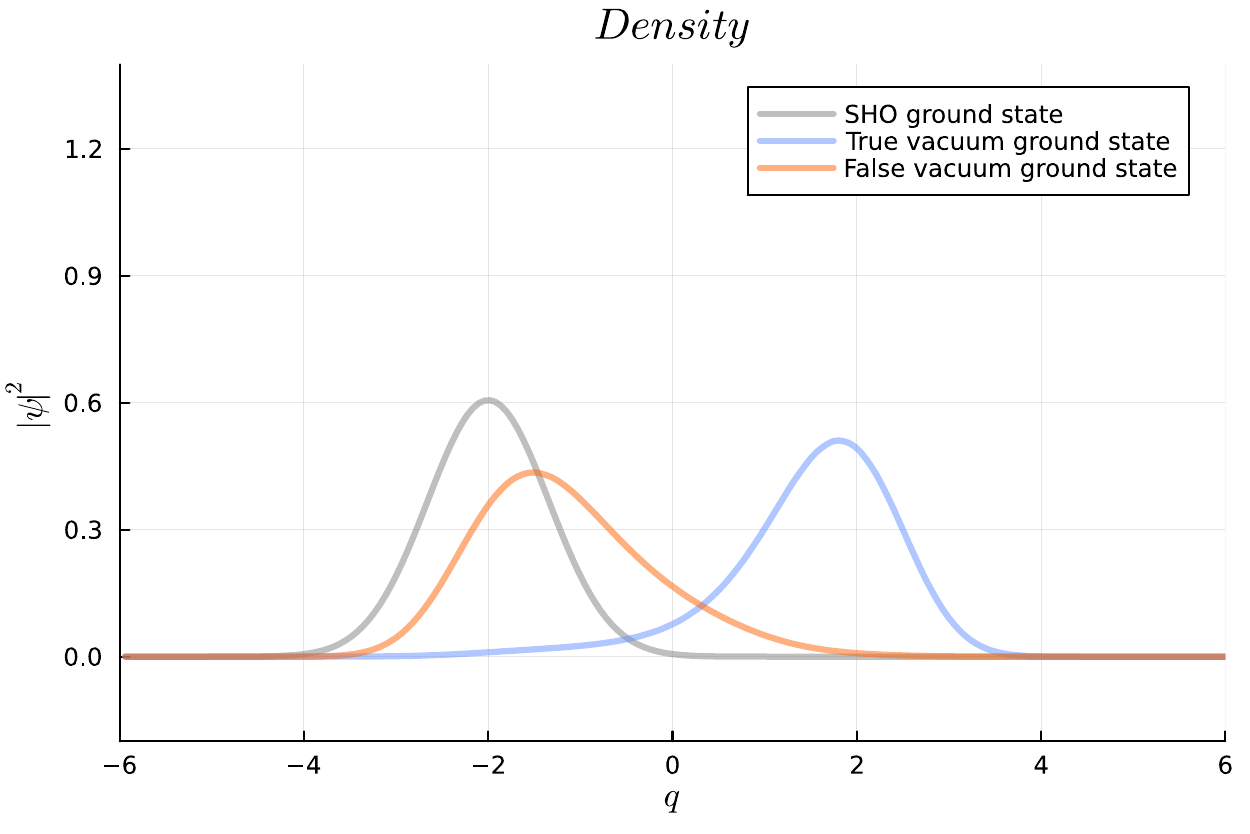}
\caption{The SHO groundstate versus the true ground state for the full potential, and the ``false vacuum ground state''.    Both of these states were found by adiabatically evolving from the SHO ground state. As might have been anticipated from the shape of the full potential compared to the SHO potential, the ``false vacuum ground state'' is shifted to the right and broadened compared to the SHO ground state. The simulation has been run for $t=300$ with a $\delta t=0.01$.}
\label{fig:poly_adiabatic}
\end{figure}

We can implement this idea using a Trotter-Suzuki decomposition of the adiabatic evolution into  steps $\delta s=1/M$, by applying the following gates to the initial qumode:
\begin{equation}
  \widetilde  {\mathcal{W}}_n ~=~ \prod _{r=1}^M \left[ e^{-i\hat p_n^2 \delta t}
e^{-i  \frac{\omega^2 }{2} (\hat q_n -\fv)^2\delta t} ~
 {\widetilde {\mathcal{V}}}^{\,r}_n  \right] ~, 
\end{equation}
where 
\begin{equation}
    \widetilde{\mathcal{V}}_n ~=~  e^{-i\scV_1 (\hat q_n) \delta s \, \delta t} ~,
\end{equation}
and where the Trotter step $\delta t$, which is chosen to be appropriately small, incorporates normal time evolution.
Note that this requires only the single non-linear gate $\widetilde{\mathcal{V}}_n $ for a qumode, which provides the ramping, and in principle one need only determine and record the correct initial qumode state once for any given potential (although in practice it is possible to simply include this procedure as part of the qumode initialisation of each QFT). 

In order for the process to remain adiabatic, the total time $t_{\rm tot}=M\delta t$ must be sufficiently long. The criterion proposed in Ref.~\cite{10.1063/1.2798382} is that the evolution satisfies 
\begin{equation}
t_{\rm tot} ~\gg ~ \frac{ 1}{ (\mathscr{E}_1-\mathscr{E}_0)_{\rm min}^2}~,
\label{eq:adiabatic_constraint}
\end{equation}
where
\( \mathscr{E}_0(t) \) and \( \mathscr{E}_1(t) \) are the eigenenergies of the ground state and first excited state at time $t$. This implies that the ground state of the full potential is more difficult to find using this method find than the ``ground state of the false vacuum''. Indeed for the former, it is reasonable to suppose that ${\mathscr{E}_1-\mathscr{E}_0}\, \approx \, 4 \varepsilon$, which is parametrically small: thus we require 
\begin{equation}
    M~\gg ~ 1/\varepsilon^2 \delta t~, 
\end{equation}
which (since we also require $\delta t \ll 1$ to minimize Trotter errors) can imply thousands of Trotter steps in order to achieve a sufficiently adiabatic evolution. On the other hand, when determining the ``false vacuum ground state'', that is adiabatically evolving to the potential with $\scV_{\rm lift}$ added, there is no near degeneracy between the ground state and the first excited state, and we have ${\mathscr{E}_1-\mathscr{E}_0}\,\approx \, 1$. Thus, in order to find the ``false vacuum ground state'' one requires only 
\begin{equation}
    M~\gg~1/\delta t~,
\end{equation} 
and adiabatic evolution through tens of Trotter steps suffices to determine the initial state in most cases.

We show an example of this procedure in Fig.~\ref{fig:poly_adiabatic} for the quartic potential of Fig.~\ref{fig:adiabatic_proc} (which is a somewhat extreme case). The curves shown are the initial SHO ground state which is centred around $\varphi=-\varphi_0$, the true ground state which is of course centred around $\varphi=\varphi_0$, found by adiabatically evolving to the full potential (requiring at least $10^4$ Trotter steps in order to remain adiabatic in this case),  and the ``false vacuum ground state''  found by adiabatically evolving to the full $+$ lift potential.

\subsection{Non-renormalisable theories: the double $\tanh (\varphi^\ell)$ well  potential}

\begin{figure}[t!]
\centering
\includegraphics[keepaspectratio, width=0.45\textwidth]{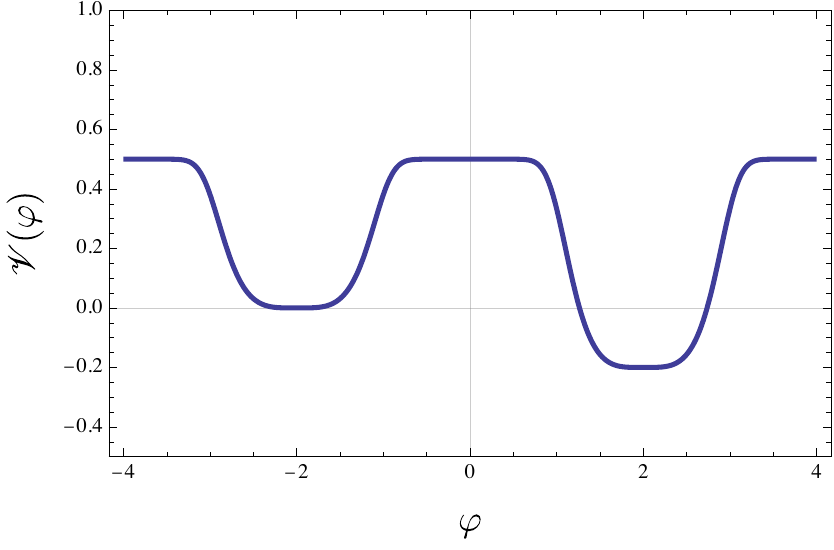}
\caption{The double $\tanh \varphi^\ell$ potential of Eq.~\eqref{eq:tanhx2_pot}  with  $\tv=-\fv=2$, $\ell = 4$ and $\varepsilon=0.2$. Potentials such as this one with $\ell>2$ give rise to ``massless'' QFTs. }
\label{fig:tanhx2_pot}
\end{figure}

It is interesting to consider alternatives to the quartic theories of Eq.~\eqref{eq:pot_poly}. 
Potentials with flat barriers are of interest because they can lead to homogeneous decay processes that are not described by the usual Callan-Coleman instanton (see for example 
\cite{JENSEN1989693} for a review and references thereof). Such theories can, for example,  undergo tunnelling through the Fubini instanton (in four dimensions), or can tunnel preferentially via thermal or stochastic processes. They can also give rise to long lived oscillations known as ``oscillons'' (\cite{PhysRevD.49.2978, PhysRevD.52.1920} -- see Ref.~\cite{Zhang:2020bec} for a recent review of the phenomenon).

A nice example of this is the following potential:
\begin{align}
\label{eq:tanhx2_pot}
{\mathscr V} (\varphi ) ~&=~
\frac{\lambda }{\ell ! }\tanh (  \mu^{-\ell}(\varphi-\fv
   )^\ell) \\
& ~~~~~~~+ ~\left(\frac{\lambda }{\ell !}+\varepsilon \right) \left( \tanh ( \mu^{-\ell} (\varphi -\tv)^\ell)-1\right)~,
\nonumber
\end{align}
which we show in Fig.~\ref{fig:tanhx2_pot} for $\ell=4$, $\mu=1$ and $\lambda = 12$. (Note that we add the $\mu$ coefficients for  dimensions.) 

These potentials would rise to non-renormalisable QFTs in four space-time dimensions, that is, QFTs that have terms of order higher than $\varphi^4$ in their potentials. (Ultimately we have four dimensional space-time in mind despite the fact that for the current study we will limit our discussion to (1+1)-dimensional QFTs).
Indeed expanding the hyperbolic functions, we see that the potential around the false minimum is approximated by 
\begin{align}
\label{eq:tanhx2_pot_approx}
{\mathscr V} (\varphi ) ~&=~
\frac{\lambda }{\ell ! }
(\varphi-\fv
   )^\ell - 
   \frac{\lambda }{3 \ell ! }
(\varphi-\fv
   )^{3\ell}~+~\ldots 
\end{align}
We thus see another interesting aspect of such potentials which is that they  encompass massless theories, that is, theories that have no $\varphi^2$ term in the potential. For example, the renormalisable terms in the $\ell=4$ example define a pure $\lambda \varphi^4$ theory which is corrected by non-renormalisable terms only at very high order, $\varphi^{12}$. One expects such ``flat potential'' theories  to behave quite differently from generic massive scalar theories.

\begin{figure}[t!]
\centering
\includegraphics[keepaspectratio, width=0.45\textwidth]{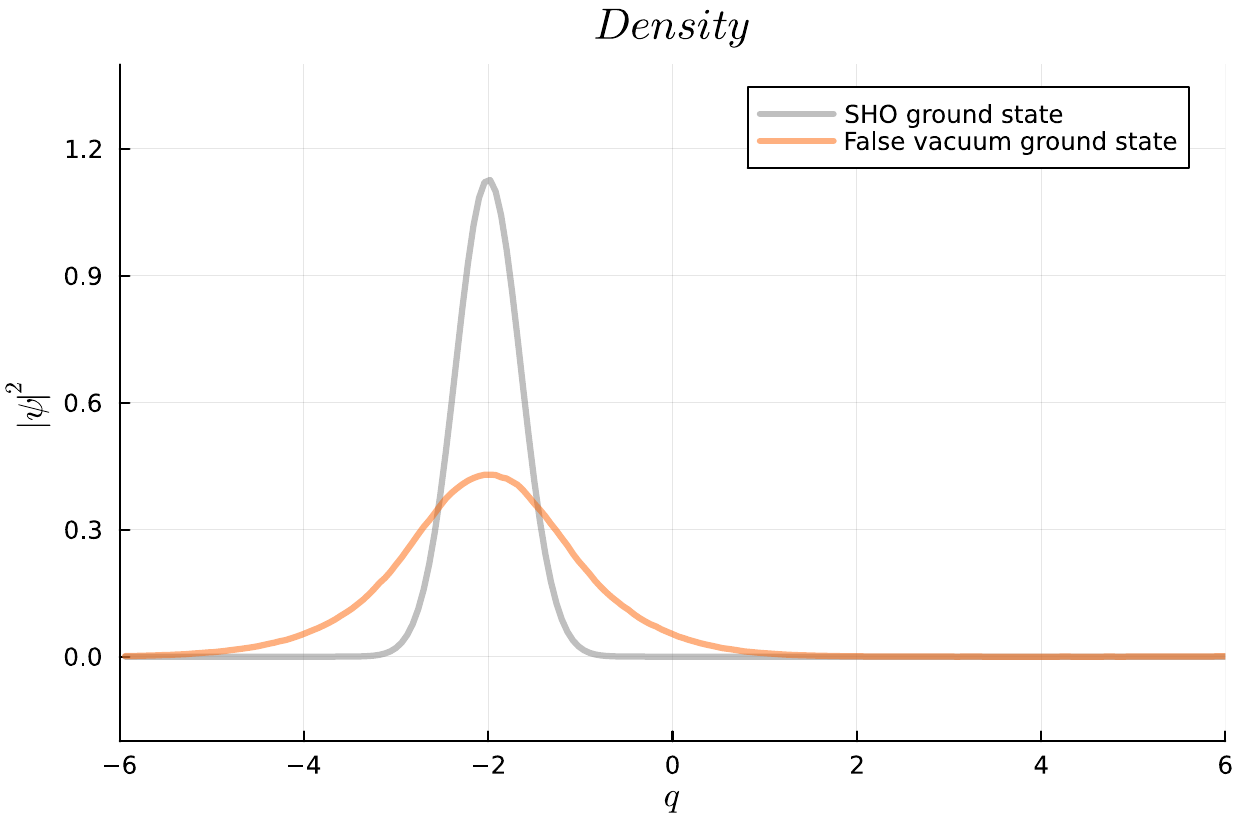}
\caption{The SHO groundstate versus the ``false vacuum ground state'' for the double $\tanh \varphi^4$ well potential of Eq.~\eqref{eq:tanhx2_pot}, with $\lambda = 12$.}
\label{fig:tanh_adiabatic}
\end{figure}

The correct ``false vacuum ground state'' can be determined adiabatically as described above, regardless of the fact that the theory is massless. Indeed applying a lift potential $\scV_{\rm lift} (\hat q_n ) $ in such theories is much simpler because one can choose a $\scV_{\rm lift}$ which simply removes the second term (which corresponds to the global minimum) from $\scV$. 
We show an example in Fig.~\ref{fig:tanh_adiabatic} for the $\ell = 4$ theory. In this massless case, there is obviously no preferred mass for the initial SHO potential, and one can simply choose its value so that the initial SHO ground state energy is comparable to that of the final ground state energy to minimise the constraint in Eq.~\eqref{eq:adiabatic_constraint}.

\subsection{Solvable non-renormalizable qumodes: the P\"oschl-Teller potential}

\begin{figure}[t!]
\centering
\includegraphics[keepaspectratio, width=0.45\textwidth]{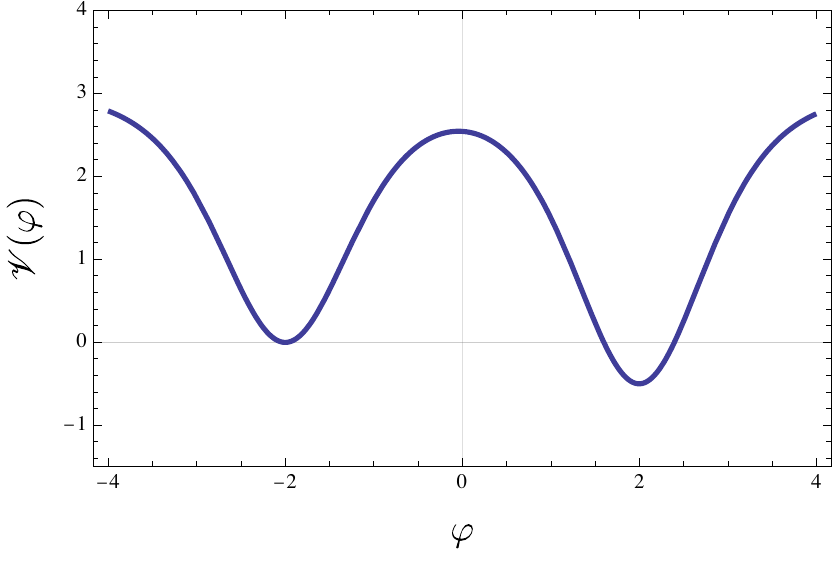}
\caption{The double P\"oschl-Teller potential of Eq.~\eqref{eq:pt_pot}  with  $\tv=-\fv=2$, $\alpha=1$, $\gamma = 2$ and $\varepsilon=0.5$.}
\label{fig:pot}
\end{figure}

It is also interesting to consider theories based on the P\"oschl-Teller (PT) potential, shown in Fig.~\ref{fig:pot}.
These theories are  outliers in quantum mechanics because they display unusual phenomena (see \cite{MARTINEZESPINOSA2024107455} for a review), and they  are solvable. This means that in classical simulations one can initialise the qumodes analytically, avoiding the need for any kind of adiabatic initialisation process described above for the quartic potentials. 

Moreover, they are of interest in the present context because the field theories that they will give rise to are also non-renormalisable.
Therefore, we will also consider the potential 
\begin{align}
\label{eq:pt_pot}
{\mathscr V} (\varphi ) ~&=~
\frac{\alpha ^2 \gamma (\gamma+1) }{2 }\tanh ^2(\alpha  (\varphi-\fv
   )) \\
& ~~~~~~~+ ~\left(\frac{\alpha ^2\gamma(\gamma+1)}{2}+\varepsilon \right) {\rm sech} ^2(\alpha  (\varphi -\tv))~.
\nonumber
\end{align}
For analytical purposes, this potential, which is essentially a double PT potential-well, can be very well approximated using the analytic methods that are applicable to the single PT potential because the two separate minima can be exponentially well separated.

Consider starting the qumode sitting in the ground state of the single PT potential,  
\begin{align}
\label{eq:PT_pot}
{\mathscr V}_{\rm PT} (\varphi ) ~&=~
\frac{\alpha ^2\gamma(\gamma+1)}{2 }\tanh ^2(\alpha  (\varphi-\fv 
   ))~.
\end{align}
  The energy eigenstates can be solved exactly by the method of ladder-operators, with the normalised ground state wavefunction being given by 
\begin{align}
\psi_n^{(0)} (q_n) ~=~ \langle q_n |   0 \rangle_- ~=~ {\cal N}_0 ~{\rm sech}^\gamma (\alpha ( q_n -\fv)) 
\end{align}
with energy given by 
\begin{equation}
    \mathscr{E}^-_0 ~=~ \frac{\alpha^2}{2} \gamma ~,
\end{equation}
and with normalisation constant 
\begin{equation}
   {\cal N}_0 ~=~  {\sqrt{\frac{\alpha}{\sqrt{\pi}}\frac{ \,\Gamma \left(\gamma +\frac{1}{2}\right)}{~\Gamma (\gamma )}}}~.
\end{equation}
The entire tower of states can be written more generally in terms of associated Legendre polynomials of the form 
\begin{align}
\psi_n^{(\nu)} (q_n) ~=~ \langle q_n | \nu \rangle ~=~ {\cal N}^{\gamma -\nu}_\gamma  P^{\gamma-\nu} _\gamma (\tanh(\alpha (q_n-\fv)) )~,
\end{align}
where $ {\cal N}^{m}_\ell  = \sqrt{{m}{\alpha }\frac{(\ell-m)!}{(\ell+m)!}}$,
with $\nu = 0\ldots \gamma -1$, and energy 
\begin{equation}
    \mathscr{E}^-_\nu ~=~ \frac{\alpha^2}{2}\left( \gamma(\gamma+1) - (\gamma - \nu)^2 \right) ~.
\end{equation}
Note that in the limit $\alpha \to 0$ the vacuum energy becomes the simple harmonic oscillator (SHO) vacuum energy of $1/2$ (which is one way to show that $\lim_{\alpha \to 0} \left[ {\rm sech}^{1/\alpha^2} (\alpha q_n) \right] \propto e^{-q_n^2 /2}$).

The same method can be applied to wavefunctions localised around the single PT well that approximates the {\it true minimum} at $q_n\approx \tv$ with its prefactor $\left(\frac{\alpha ^2\gamma(\gamma+1)}{2}+\varepsilon \right)$. 
There is an interesting choice of $\varepsilon$ for this potential, namely  
$\varepsilon = \alpha^2 (\gamma +1)$. For this choice the prefactor becomes $\frac{\alpha ^2(\gamma+1)(\gamma+2)}{2}$,  
and the energies in the PT well around the true minimum become  
\begin{align}
    \mathscr{E}^+_\nu ~&=~ \frac{\alpha^2}{2}\left( (\gamma+1)(\gamma+2) - (\gamma + 1  - \nu)^2  \right)~- \varepsilon ~\nonumber \\
    ~&=~ \frac{\alpha^2}{2}\left( \gamma (\gamma+1) - (\gamma + 1  - \nu)^2  \right)~~.
\end{align}
We thus find that $\mathscr{E}^-_0 = \mathscr{E}^+_1$ for this choice of $\varepsilon$, so that the metastable ground state around the minimum at $q_n=\fv$ has the same energy as the first excited state around the true minimum at $q_n=\tv$. In such a situation, the ground state of the PT well at $\varphi = \fv$ is actually a superposition of just the two true energy eigenstates of the whole system that have nearly degenerate energies. If we denote the towers of states in each potential well as $|\nu\rangle _\pm$, then we can identify these first two excited energy eigenstates (up to irrelevant phases) as
\begin{align}
    |{\mathscr{E}}_1 \rangle ~&=~ 
    \cos \theta  |0\rangle_-  +
   \sin \theta  |1\rangle_+ \nonumber \\
    |{\mathscr{E}}'_1\rangle ~&=~ - \sin \theta  |0\rangle_- +
    \cos \theta |1\rangle_+ ~,
\end{align}
where ${\mathscr{E}}'_1 \gtrsim {\mathscr{E}}_1\approx {\mathscr{E}}^-_0 = {\mathscr{E}}^+_1$, for some angle $\theta$. (As the potential is not symmetric we cannot use parity to fix the coefficients.) Conversely if we start the qumode in the ground state $|0\rangle_-$ of the metastable minimum, it will evolve as 
\begin{align}
\label{eq:evolve}
    |\psi _n(t) \rangle ~&=~ 
    \cos (\Delta\mathscr{E}\,t /2 ) \, |0\rangle_- \\
    & ~~~+~
    i  \sin (\Delta\mathscr{E}\,t /2 ) \, \left( \cos 2\theta  |0\rangle_- + \sin 2\theta|1\rangle_+   \right)  ~\nonumber 
\end{align}
where $\Delta\mathscr{E} = {\mathscr{E}}'_1-{\mathscr{E}}_1$. Now the energy splitting between these two states is known to be generated non-perturbatively by the ($1+0$ dimensional) tunnelling instanton, and is given by 
\begin{equation}
    \Delta\mathscr{E} ~\sim ~  {\mathscr{E}}_1 e^{-S_1  }
\end{equation}
where the instanton action is 
\begin{equation}
    S_1 ~\approx~  \int_{\fv} ^{\varphi_{\rm esc} }\sqrt{2{\mathscr{V}(q_n)}}~dq_n~. 
    \label{eq:S1}
\end{equation} Given  Eq.~\eqref{eq:evolve}, for this particular choice of $\varepsilon $ we expect the state to oscillate back and forth between the initial state and the first excited state around the true minimum in a manner determined by the mixing angle $\theta$ and the instanton action $S_1$.

\begin{figure}[t!]
\centering
\includegraphics[keepaspectratio, width=0.45\textwidth]{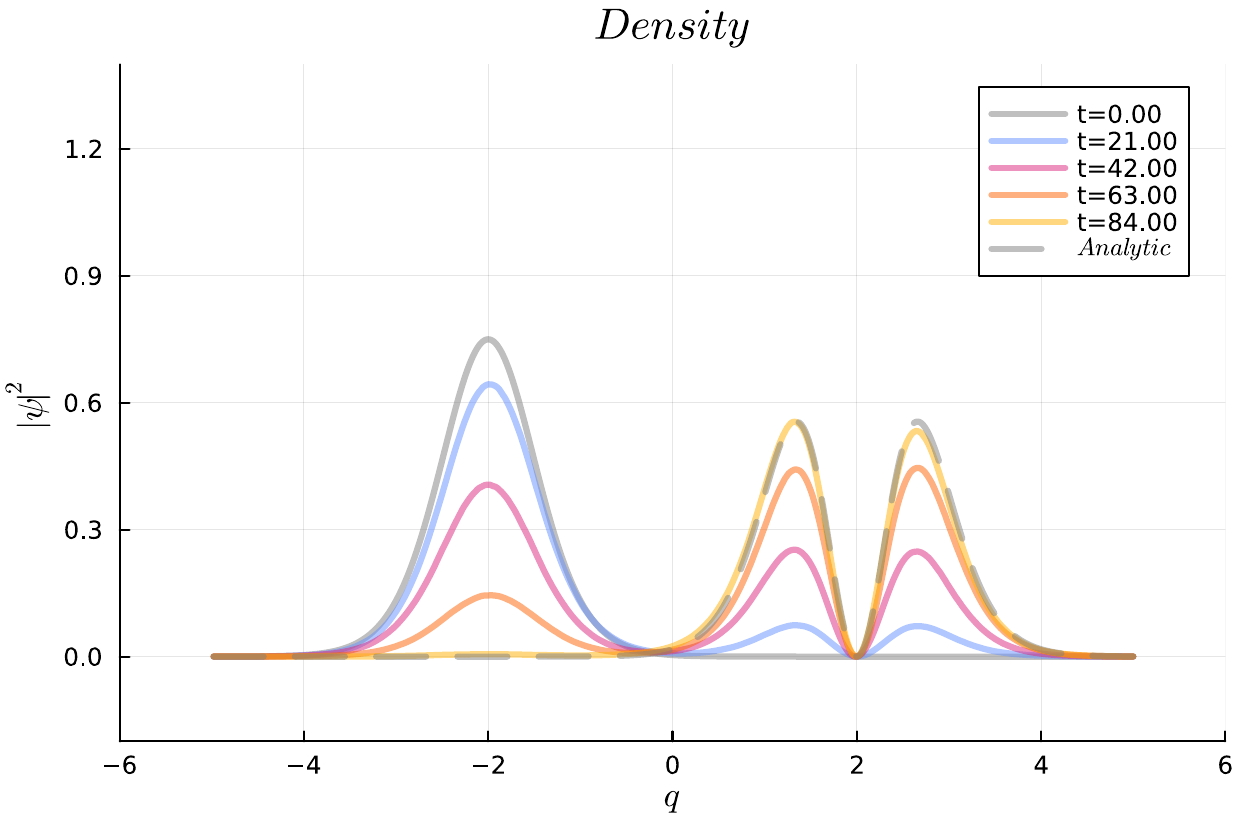}
\caption{Trotter-Suzuki evolution of a single qumode under the P\"oschl-Teller potential in Eq.~\eqref{eq:pt_pot} with $\alpha =1$, $\tv=-\fv=2$, $\gamma =2$ and $\varepsilon = 3$. The dashed line represents the first excited state, $\langle q |1\rangle_+ \sim P^1_2(\tanh q)$ of the PT potential-well at $q_n=2$.}
\label{fig:wobble}
\end{figure}

The numerical evolution is shown in Fig.~\ref{fig:wobble}, which clearly conforms to this expectation. Thus this figure constitutes a direct measurement of the instanton in quantum mechanical tunnelling using a single qumode. In fact, since the initial $|0\rangle_-$ state oscillates to pure $|1\rangle_+$, we can conclude that $\theta = \pi/4$, so that, despite the potential being asymmetric, the energy eigenstates consist of the same odd/even linear superpositions  $|0\rangle_- \pm |1\rangle_+$ that would be found for the symmetric potential. In addition we determine the energy splitting $\Delta{\mathscr{E}}$ that the instanton induces in this pair of states to be $\Delta{\mathscr{E}} = \pi/21$.

\subsection{The qumode lattice}

Now that we understand how to construct and initialise a single qumode, and hence reproduce their local quantum mechanics, let us tie them together into a one-dimensional lattice to reproduce QFT. First of course we must augment the operator $\widetilde{\mathcal{U}}_n$ with the contribution to the Hamiltonian coming from the diagonal terms in Eq.\eqref{eq:qumode_ham}. That is we have 
\begin{equation}
{\mathcal{U}}_n(\delta t) ~=~ e^{-i\hat p_n^2 \delta t}
    e^{-i\left( \hat q_n ^2 + \mathscr{V}(\hat q_n) \right)\delta t} ~,
\end{equation}
with overall 
\begin{equation}\label{eqn:diag}
{\mathcal{U}}_{\rm diag}(\delta t) ~=~ \prod_{n=1}^N {\mathcal{U}}_n(\delta t)~. 
\end{equation}
Next we come to the non-diagonal contribution to the QFT evolution which arises from the cross-terms in the discretised Hamiltonian of Eq.~\eqref{eq:disc_ham}; namely we are required to also consider the Hubbard-like evolution operator 
\begin{equation}\label{eq:qhop}
\mathcal{U}_\textrm{hop}(\delta t) ~=~ \prod_{n=1}^N  e^{i \, \hat{q}\,_{{n+1}} \,\hat{q}_n \, \delta t}~
\end{equation}
acting on the lattice. 
Thus we have in the first order Trotter-Suzuki decomposition a total evolution operator of the form 
\begin{equation}
    {\cal U}(\delta t)  ~=~  \mathcal{U}_\textrm{diag}(\delta t)~\mathcal{U}_\textrm{hop}(\delta t)
\end{equation}
with the exact evolution operator over time $t$ being expressed as 
\begin{align}
    \cU_{\rm exact}(t)  ~&=~
       \left[ \cU (\delta t) \right] ^{t/\delta t} + \cO(\delta t^2)
\end{align}
 It is possible to make improvements (at the expense of adding more gates) which make the Trotter-Suzuki decomposition effectively higher order. In this study we will use instead the second order decomposition 
 \begin{equation}
    {\cal U}(\delta t)  ~=~  \mathcal{U}_\textrm{diag}(\delta t/2)~\mathcal{U}_\textrm{hop}(\delta t/2) ~\mathcal{U}^T_\textrm{hop}(\delta t/2) ~\mathcal{U}_\textrm{diag}(\delta t/2)~,
\end{equation}
which gives 
\begin{align}
    \cU_{\rm exact}(t)  ~&=~
       \left[ \cU (\delta t) \right] ^{t/\delta t} + \cO(\delta t^3)~.
\end{align}
Note that there is a trade-off to be made. It is computationally twice as expensive to construct the gates required for such a second order evolution. However, this is a one-time calculation that must be performed at the beginning of the evolution, and meanwhile the number of time steps required to achieve the same precision in the evolution decreases by a power of $2/3$. For very entangled systems it is likely to be favourable to reduce the number of Trotter steps that have to be evaluated.

Let us now turn to the practical implementation of a qumode lattice on a tensor network (TN)~\cite{itensor}. Using the matrix product state (MPS)~\cite{PhysRevB.55.2164, perezgarcia2007} and matrix product operator (MPO)~\cite{pirvu2010matrix} ans\"atze, we simulate the second-order Trotterised real-time evolution of (1+1)-dimensional scalar field theory using the time-evolving block decimation (TEBD) algorithm~\cite{PhysRevLett.91.147902, PhysRevLett.93.040502}, which proceeds in three main stages. 

MPS are a class of variational ans\"atze that efficiently capture the low-entanglement structure of one-dimensional quantum many-body systems. Formally, an MPS expresses the wavefunction of a system with N sites as a chain of rank-3 tensors $\{A^{[n]}\}$, such that the total state reads
\begin{equation}
|\psi\rangle = \sum_{\{i_n\}} A^{[1]i_1} A^{[2]i_2} \cdots A^{[N]i_N} |i_1 i_2 \ldots i_N\rangle,
\end{equation}
where each tensor $A^{[n]i_n}$ carries a physical index $i_n$ of dimension $d$, and auxiliary bond indices of dimension $\chi$, the so-called bond dimension. The bond dimension controls the amount of bipartite entanglement that can be encoded across a cut in the system and thus determines the expressiveness of the ansatz: low-$\chi$ MPS efficiently approximate ground and low-energy states of gapped local Hamiltonians due to the area law for entanglement entropy \cite{}. In the context of time evolution, such as with the TEBD algorithm, the MPS structure allows for efficient application of local gates and entanglement truncation via singular value decomposition, thereby rendering the simulation of real-time dynamics tractable even when the full Hilbert space grows exponentially with system size.

Thus, one must construct a MPS representation of the initial state by factorising the exponentially large Hilbert space of the system into a product state of local Gaussian wavepackets. The MPS is then evolved forward in time by applying one and two-site Trotterised  gate operations are implemented as matrix operations of the form shown in Equations~\eqref{eqn:diag} and \eqref{eq:qhop}, respectively. Finally, we perform an entanglement truncation on the system. This is done by truncation the MPS bond-dimension via singular-value decompositions.

Then, the TEBD algorithm is a real-time simulation technique designed for one-dimensional quantum many-body systems with local (typically nearest-neighbour) interactions, and is especially efficient when the quantum state is represented as an MPS. The essential idea of TEBD is to approximate the full time-evolution operator $U(t) = e^{-iHt}$ by Trotter-Suzuki decomposition, splitting H into a sum of terms that act on disjoint pairs of sites, typically of the form
\begin{equation}
H = \sum_{\text{even } n} h_{n,n+1} + \sum_{\text{odd } n} h_{n,n+1},
\end{equation}
so that exponentials of each commuting set can be applied in parallel. A second-order decomposition, for example, proceeds via

$$
 U(t)   \approx  \left[\prod_{\text{odd } n} \\
 e^{-i {\bf h} \delta t/2} \prod_{\text{even } n} e^{-i {\bf h} \delta t} \prod_{\text{odd } n} e^{-i {\bf h} \delta t/2} \right]^{t/\delta t},
 $$
where ${\bf h} \equiv h_{n,n+1}$ and the operators $e^{-i h_{n,n+1} \delta t}$ are two-site unitaries that act locally on pairs of adjacent sites in the MPS. This structure requires the Hamiltonian to be at most nearest-neighbour coupled, or at least decomposable into local gates.

Each gate is applied sequentially by contracting it into the relevant MPS tensors, followed by a singular value decomposition (SVD) to restore the MPS form. The SVD naturally produces a Schmidt decomposition across the bond, and a truncation is applied by discarding singular values below a fixed threshold or keeping only the top $\chi$ singular values. This is what allows TEBD to dynamically control entanglement growth during evolution, ensuring computational feasibility as long as the entanglement entropy remains moderate.

The TEBD algorithm is particularly well-suited for systems with short-range interactions and limited entanglement, such as gapped ground states and mildly entangled real-time dynamics. In the present context, the qumode lattice Hamiltonian consists of a sum of on-site (diagonal) terms and nearest-neighbour interaction terms, rendering it directly compatible with TEBD's gate-splitting structure. The algorithm proceeds alternately applying single-site gates (on-site potentials and kinetic energy terms) and two-site gates (spatial derivative couplings) to simulate the full dynamics of the field theory.

\section{Initial state preparation}

\label{subsec:stateprep}

We now turn to the initial state preparation for the entire qumode lattice. 
We have seen that it is possible to  determine the ``false vacuum ground state'' of a single uncoupled anharmonic qumode using the adiabatic theorem. However the ground state of the QFT is not simply the product of ground states of the uncoupled anharmonic qumode oscillators. As discussed in Ref.~\cite{Abel:2025zxb} the QFT ground state {\it approaches} the product of uncoupled qumode ground states for modes in the large mass and low momentum limit, but generally it will involve multi-qumode entanglement and hence long-range correlations. This distinction could be ignored in Ref.~\cite{Abel:2025zxb} because that work  was concerned only with scattering processes, so it could accordingly focus on scattering of low momentum wave-packets. Unfortunately for tunnelling processes the distinction is important, because the long-range fluctuations that seed the process are absent from the product of qumode groundstates. In other words, quantum tunnelling could occur on the lattice spontaneously even if it is initialised as the product of qumode ground states, but it could only be a much rarer coherent transition of the whole lattice rather than a localised bubble. 

In principle one could initialise the entire lattice using the same adiabatic evolution procedure that we used for a single qumode in Sec.~\ref{subsec:adiabat}. This would entail starting the system in an MPS which consists of the product of qumodes all in their ``false vacuum ground states''  determined as described in \ref{subsec:adiabat}. Then we would act on the MPS with the following set of $M$ MPOs:
\begin{equation}
  \widetilde  {\mathcal{W}} ~=~ \bigotimes _{r=1}^M \prod_{n=1}^N \left[ e^{-i\hat p_n^2 \delta t}
e^{-i  (\scV(\hat q_n)+\scV_{\rm lift}(\hat q_n)) \delta t } ~
 {\widetilde {\mathcal{V}}}^{\,r}_n  \right] ~, 
\end{equation}
where 
\begin{equation}
    \widetilde{\mathcal{V}}_{n<N} ~=~  e^{-i ( \hat q_n^2 - \hat q_n \hat q_{n+1})  \delta s \, \delta t} ~,
\end{equation}
and where again we have $\delta s=1/M$ and the small Trotter step $\delta t$ incorporating normal time evolution. As we do not impose periodic boundary conditions, we can take $\widetilde{\mathcal{V}}_{N}=1$. 

The principle is, of course, exactly the same as for the determination of the ``false vacuum ground state'' of a single qumode. That is, beginning in an MPS which is a product of qumode oscillator ground states,  and adiabatically turning on the space-derivative couplings, which include the hopping terms, should result in the system reaching the ``false vacuum ground state'' of the QFT. Note, for this purpose that we must add $\scV_{\rm lift}$ to the potential to ensure that the system does not tunnel during the adiabatic evolution while it is being prepared. Note also that one could combine everything into a single adiabatic procedure, starting in an MPS which is a product of SHO ground states centred at $\fv$, and adiabatically turning on both $\scV_1$ and the space-derivative terms at the same time. 

Although appealing from a physics perspective, this kind of procedure has a significant drawback. On a tensor network it would be exceedingly lengthy to implement, because the evolution of the system has to remain adiabatic. Indeed the accurate adiabatic preparation of a single qumode required thousands of Trotter steps, and therefore a currently feasible QFT vacuum preparation by adiabatic evolution would be subject to errors because of loss of adiabaticity. 

In this section we present two alternative methods for QFT vacuum preparation. The first is a compromise option: namely we work in the ground state corresponding to a product of uncoupled anharmonic oscillators, and then model the effect of the missing hopping term by sampling. This is an efficient method which is useful for studying phase transitions across the whole system. 
 
The second method is a no-compromise option that requires a separate stage of preparation: namely we find the exact QFT vacuum by imaginary-time TEBD on our qumode lattice. As we shall see this yields the complete QFT ``false vacuum ground state''. 

\subsection{Approximating the ground state by sampling}
\label{subsec:sampling}

Let us consider the first method, to approximate the QFT ground state by sampling. For this method one generates an ensemble (labelled by $s$) of initial position  and momentum fluctuations, $\{\Delta q_n^{(s)}, \Delta p_n^{(s)}\}$, whose distributions incorporate those quantum fluctuations and correlations of the  QFT ground state which are missing when we simply start the system as an MPS which is a product of  quantum anharmonic oscillators. 

It is straightforward to derive the fluctuations required to simulate the false vacuum of the QFT by sampling a classical distribution. To see how this works, first suppose that there were no quantum fluctuations at all in the system, and that we wished to generate {\it all} of the fluctuations by sampling. In that case one would proceed as follows. The fluctuations that we wish to  reproduce are encompassed by the position and momentum covariance matrices of the QFT vacuum, $C_q$ and $C_p$, which are $N \times N$ matrices characterising the correlations between sites:
\begin{align}
C_q &= \langle \hat{q} \hat{q}^T \rangle - \langle \hat{q} \rangle \langle \hat{q}^T \rangle, \\
C_p &= \langle \hat{p} \hat{p}^T \rangle - \langle \hat{p} \rangle \langle \hat{p}^T \rangle,
\end{align}
where $\hat{q} = (\hat q_1, \hat q_2, \ldots, \hat q_N)^T$ and similarly for $\hat{p}$. To find a suitable ensemble of $q$ and $p$ values that reproduces these covariance matrices, we Cholesky decompose them as follows:
\begin{align}
C_q &~=~ L_q L_q^{T}, \nonumber \\
C_p &~=~ L_p L_p^{T},
\end{align}
where $L_q$ and $L_p$ are lower-triangular matrices. It is straightforward to show that an ensemble of displacements and momenta of the following form will reproduce the same covariance matrices (where $s$ simply labels the members of the ensemble):
\begin{align}
{q}^{(s)} ~&=~ {q}_{\text{mean}} + L_q {\eta}^{(s)}_q~,\nonumber \\
{p}^{(s)} ~&=~ {p}_{\text{mean}} + L_p {\eta}^{(s)}_p~.
\end{align}
Here $q_{\text{mean}}$ and $p_{\text{mean}}$ are the mean initial displacement and momentum vectors (e.g., $q_{\text{mean}} = \fv$ if starting from the ground state in the false vacuum), and  $\eta^{(s)}_{q,p}$ are standard normal random vectors,
\begin{align}
{\eta}_q, {\eta}_p \sim \mathcal{N}(\bm{0}, \mathbf{I}_N)~,
\end{align}
where ${\eta}^{(s)}_q, {\eta}^{(s)}_p \in \mathbb{R}^N$.

The lattice would then be initialised with position and momentum displacements drawn from this ensemble:
\begin{itemize}
\item Positions: \(\{q_j^{(s)}\}_{j=1}^N\)
\item Momenta: \(\{p_j^{(s)}\}_{j=1}^N\)
\end{itemize}
and the dynamical simulation can then proceed as normal. 

In the current context, as we are interested mainly in missing long-range correlations, it will be sufficient to use the free-field SHO approximation around the false vacuum. This is given by~\cite{Abel:2025zxb}
\begin{equation}
    \langle {\bf q}|\Psi \rangle ~ \sim ~ \prod _\alpha \left( {\omega_\alpha}^{-\frac{1}{4}} ~e^{  - \omega_\alpha |q_\alpha |^2 } \right)  ~,
 \end{equation}
where the frequencies are given by Eq.~\eqref{eqn:omegas}.
With this approximation the computation is considerably simplified. It is simple to show that the covariance matrices are  
\begin{align}
C_{q,nm} &~=~ \sum_\alpha \frac{1}{2\omega_\alpha } U^\dagger_{n\alpha } U_{\alpha m }  \\
C_{p,nm} &~=~ \sum_\alpha \frac{\omega_\alpha}{2 } U^\dagger_{n\alpha } U_{\alpha m } ~,
\end{align}
where $U_{\alpha m}$ is as given in Eq.~\eqref{eq:unitaries}, and one can then directly identify 
\begin{equation}
    L_{q,n\alpha}  ~=~ \frac{1}{\sqrt{2\omega_\alpha}} U^\dagger_{n\alpha } ~;~~ L_{p,n\alpha}  ~=~ \sqrt{\frac{\omega_\alpha}{2}}\,  U^\dagger_{n\alpha }~.
\end{equation}
There is only one subtlety in the decomposition, which is the fact that $q_n$ and $p_n$ are real. Focussing on the position displacements for the moment, one now has 
\begin{align}
{q}^{(s)} ~=~ \fv + \sum _{\alpha = 0} ^{N-1}\frac{1}{\sqrt{2N\omega_\alpha}} e^{-2\pi i n \alpha }  {\eta}^{(s)}_{q,\alpha}~
, \end{align}
which implies that we must set $\eta^{(s)} _{q,\alpha}  = \eta ^{(s)}_{q,N-\alpha} $, or equivalently, restoring the lattice  index $n$, and selecting from two sets of standard normal vectors $\eta$ and $\eta'$, we have    
\begin{align}
{q}_n ~=~ \fv + \sum _{\alpha = 0} ^{N-1}\frac{1}{\sqrt{2N\omega_\alpha}} \, \cos (2\pi n \alpha/N )~  {\eta}_{n,\alpha}~
, \end{align}
and likewise implement initial momenta  of the form  
\begin{align}
p_n ~=~ \sum _{\alpha = 0} ^{N-1}\sqrt{\frac{\omega_\alpha}{2N}} \, \cos (2\pi n \alpha /N )~  {\eta}'_{n,\alpha}~
. \end{align}
(This constraint is equivalent to requiring that the fluctuations conserve momentum in the system as a whole.) 
This is the final expression for generating {\it all} the initial fluctuations entirely by random sampling of a classical ensemble. 

This is not quite what we want for the case at hand, because as we have already discussed the {\it low-momentum} fluctuations are all present and correct even when we start the system in an MPS which is just the product of qumode ``false vacuum ground states''. What we need to add to this initial system are just those fluctuations of the QFT vacuum which are missing.

To see how we can identify them, let us first understand a bit better the overall structure of the qumode lattice fluctuations by considering what happens to our result above when there is uniform mode variance, i.e. $\omega_\alpha =\omega$. In this case, the fluctuations in both $q$ and $p$ are proportional to 
$\frac{1}{\sqrt{N} }\sum_\alpha \cos(2\pi n \alpha ) \eta _{q,\alpha }$. This 
is essentially the Fourier transform of a white noise vector with independent and identically distributed \(\mathcal{N}(0,1)\) entries. It is a straightforward exercise to show that the variance of this sum leads to $\Delta q_n ~=~ 1/2\omega$ and $\Delta p_n ~=~ \omega/2$. This variance in the individual qumode quadrature variables is exactly equivalent to the variance of the SHO ground state: in other words, if $\omega_\alpha$ were degenerate then the QFT vacuum would be the same as the product of qumode ground states! This is of course entirely in accord with the discussion of Ref.~\cite{Abel:2025zxb} which concluded the two vacua are essentially equivalent if one is concerned with scattering of low-momentum (specifically non-relativistic) modes for which $\omega_\alpha \approx m$.

Thus we see that the required covariance matrices encompass the difference between the full contribution and the degenerate contribution from the product of qumode groundstates, namely 
\begin{align}
C_{q,nm} &~=~ \sum_\alpha \frac{1}{2}\left(\frac{1}{\omega_\alpha} -\frac{1}{\omega} \right)  ~U^\dagger_{n\alpha } U_{\alpha m }  \\
C_{p,nm} &~=~ \sum_\alpha \frac{1}{2 } \left( \omega_\alpha-\omega\right) ~U^\dagger_{n\alpha } U_{\alpha m } ~.
\end{align}
These covariance matrices can be generated by fluctuations around the product of qumode ``false vacuum ground states'' of the following form:  
\begin{align}
\Delta{q}_{n} ~&=~ \sum _{\alpha = 0} ^{N-1}\frac{1}{\sqrt{2N}}\sqrt{\left(\frac{1}{\omega} - \frac{1}{\omega_\alpha}\right)} ~ \cos (2\pi n \alpha/N )~  {\eta}_{n,\alpha}~
, \nonumber \\
\Delta p_{n} ~&=~ \sum _{\alpha = 0} ^{N-1}\sqrt{\frac{\omega_\alpha-\omega }{2N}} \, \cos (2\pi n \alpha /N )~  {\eta}'_{n,\alpha}~
. \label{eq:fluck}\end{align}

Thus as anticipated such fluctuations die away for the  low-momentum non-relativistic modes in the spectrum, which as per Eq.~\eqref{eqn:omegas} satisfy $\alpha /aN \ll \omega$. In the infinite lattice limit, these modes do not contribute to qumode fluctuations, which as mentioned is in accord with the intuition that the QFT vacuum coincides with the product of qumode ground states for the low momentum modes. By contrast, the highly relativistic modes have $\omega_\alpha \approx k_\alpha = 2\pi\alpha /aN $, and the weighted sum retains correlations. It is straightforward to show that the contributions from these relativistic modes to fluctuations in the $q_n$ grow slowly with the size of the lattice, as $\Delta q_n \sim \sqrt{\log N} $.

The upshot is that we can mimick the required long-range correlations by setting the qumodes in their ``false vacuum ground states'' and then augmenting them by random sampling the classical distributions of position and momentum fluctuations presented in Eq.~\eqref{eq:fluck}, where the angular frequencies $\omega$ and $\omega_\alpha$ are found from Eqs.~\eqref{eq:pot_poly} and 
\eqref{eqn:omegas} respectively. (The shift $\Delta p$ is of course implemented by acting on the wave function of each qumode with the operator $e^{i\Delta p_n \, \hat q_n }$.)

\subsection{Imaginary-time TEBD determination of the initial state}

\label{subsec:iTEBD}

We now discuss the second option for determination of the QFT vacuum. As mentioned this option requires no compromises and is a genuine QFT initial state preparation for vacuum tunnelling. The method works by Hamiltonian evolution of the QFT through imaginary time using TEBD following the techniques discussed in Refs.~\cite{vidal,schollwock2011density,Orus:2013kga}. That is we begin with the system as a product of SHO ground states, in the lifted false vacuum of Figure \ref{fig:adiabatic_proc}. We then perform the same Trotterized Hamiltonian evolution that we would perform for the simulation itself, but through imaginary time. The MPS will of course immediately begin to attenuate because such an evolution is non-unitary. Therefore after each Trotter step it is  renormalized by scaling. The crucial point that makes the trick work is that the rate of attenuation is proportional to the energy, and over many Trotter steps the lowest energy eigenstate dominates. Thus the system is expected to converge quite rapidly to the desired QFT false vacuum state.

The great advantage of the method is precisely that it is not physical. There is no concern about for example maintaining adiabaticity, and we do not need to worry if the system gets artificially ``kicked'' by relatively large Trotter steps, because ultimately all these perturbations will be attenuated away. 

In order to test whether the mechanism has worked we can make measurements of the same covariance matrix that we discussed in the previous section. 
That is we determine 
\begin{align}
\label{eq:Cnmagain}
    C(k_\alpha )~=~ \sum _{n,m}  C_{nm} U_{\alpha n } U^\dagger_{m\alpha }~,
\end{align}
 by measuring the matrix 
\begin{align} 
C_{nm} ~=~ \langle (\hat q _n - \langle q_n\rangle ) (\hat q _m - \langle q_m\rangle )  \rangle 
\end{align}
on the lattice, where $\langle q_n\rangle \approx \fv$. 

Ideally one would expect to find 
\begin{align} 
C(k_\alpha )~=~ \frac{1}{2\omega_\alpha}~,
\end{align} 
as we have already discussed. 

However, a fit using Eq.~\eqref{eqn:omegas} with the exact mass deduced from the potential in $\omega_\alpha$ is not good. It turns out that for an accurate fit we need two parameters in $C(k_\alpha)$. One of these parameters is an {\it effective mass} which we shall call $m$ which is inserted into the dispersion relation Eq.~\eqref{eqn:omegas} in place of the ideal mass deduced from the potential which would have $m=\omega$. The second parameter is a constant $C_0$ which corresponds to an effective extra $k_\alpha$ independent on-site variance.  Thus we fit 
\begin{align} 
\label{eq:CK}
C(k_\alpha)~=~ \frac{1}{2\omega_\alpha (m)}~+ ~C_0~.
\end{align} 

\begin{figure}[t!]
\centering
\includegraphics[keepaspectratio, width=0.45\textwidth]{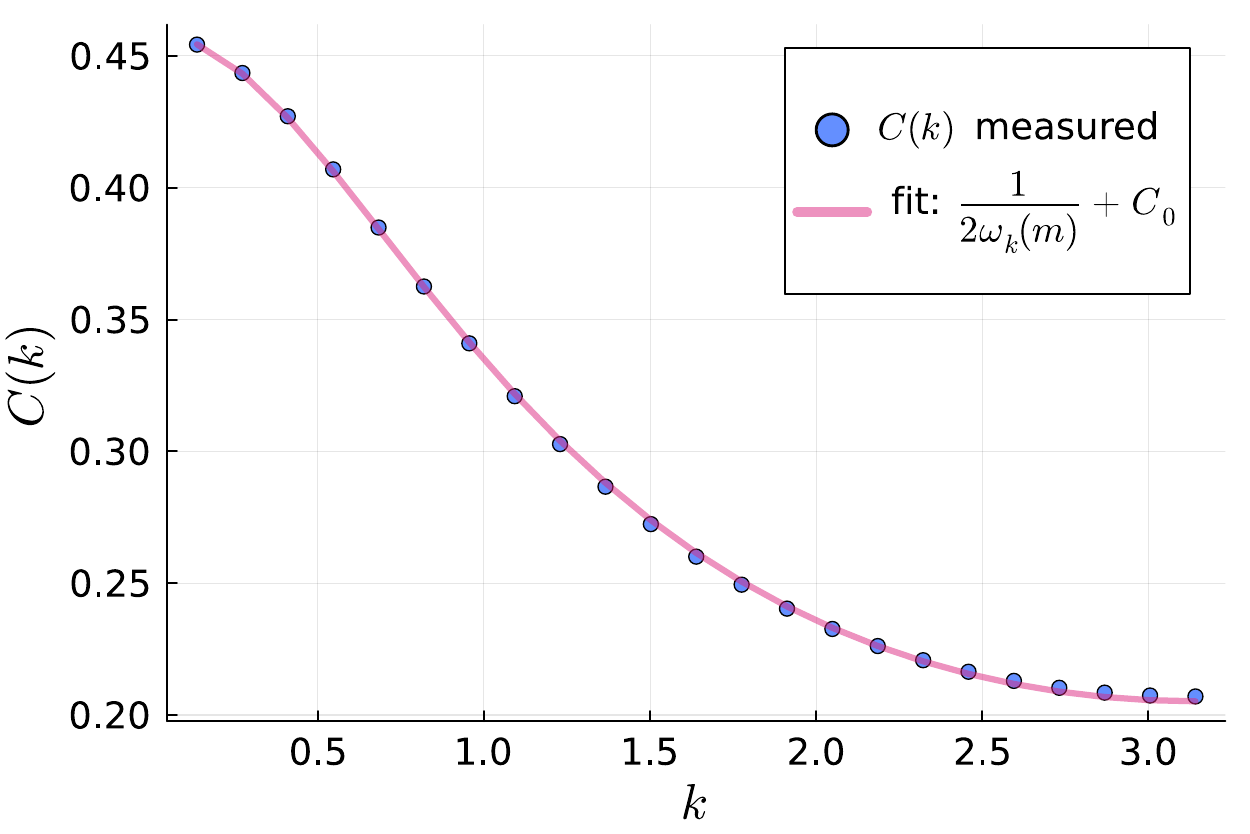}
\caption{Comparison of the measured covariance matrix of the QFT vacuum versus the theoretical expressions with fit parameters $m=1$ and $C_0=-0.016$, for the SHO potential centred around the origin with $\omega=1$.}
\label{fig:Sk_SHO}
\end{figure}

\begin{figure}[t!]
\centering
\includegraphics[keepaspectratio, width=0.45\textwidth]{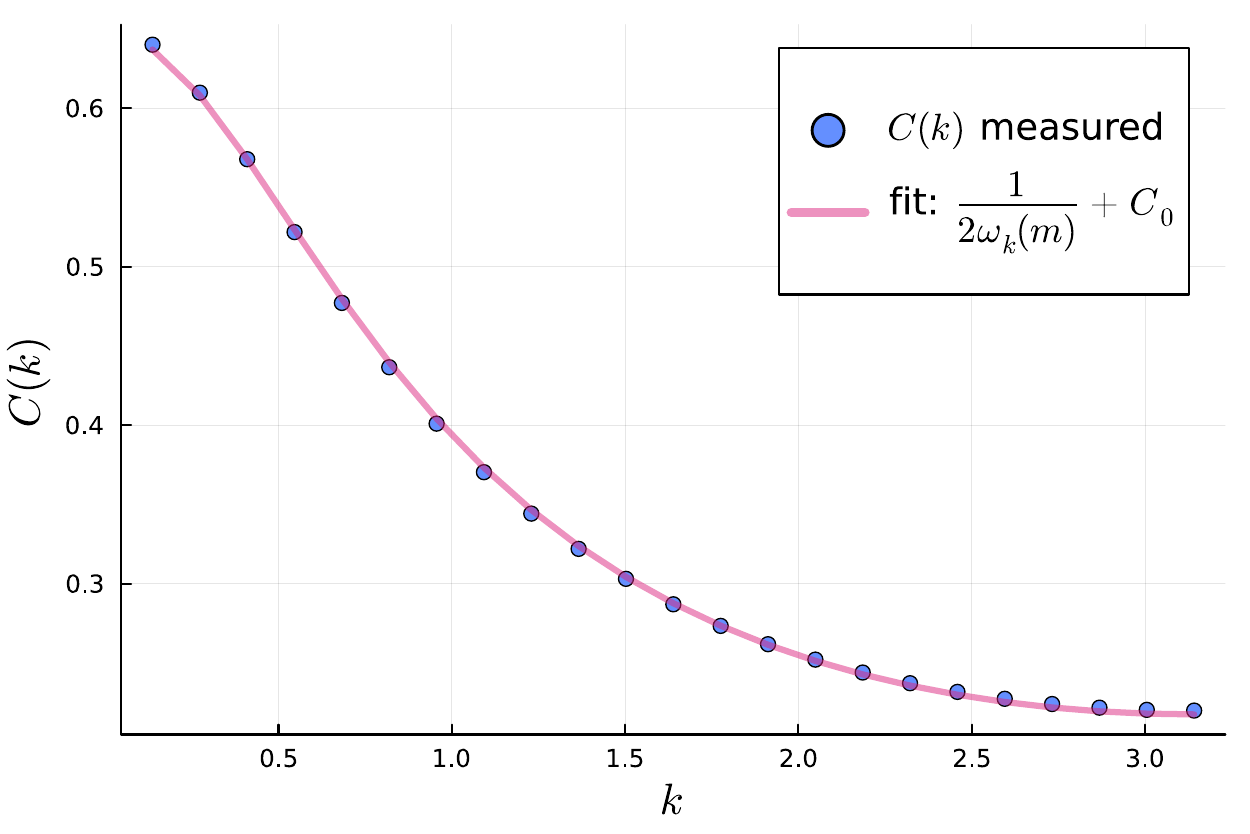}
\caption{Comparison of the covariance matrix of the QFT vacuum versus the theoretical one, for the PT potential in the false vacuum at $\varphi_0 = -2$. Here the best fit has $m=0.72$ and $C_0=-0.016$. }
\label{fig:Sk_PT}
\end{figure}

We will investigate the meaning of the two  parameters $m$ and $C_0$ shortly. For the moment let us present the fits, which are shown in Figs.~\ref{fig:Sk_SHO} and \ref{fig:Sk_PT} for the unit mass SHO potential with 
\begin{equation}
\scV(\varphi ) ~=~ \frac{1 }{2} (\varphi -\fv)^2 ~,
\label{eq:shopot}
\end{equation}
and for the PT potential of Eq.~\eqref{eq:pt_pot} respectively,  lifted and with parameters ($\alpha = 1, \gamma=\frac{\sqrt{5} - 1}{2}$) chosen such that it also has $\omega =1$. In this case we are taking 20 quadrature points for our qumodes (\ie they are approximated by a local Hilbert space of dimension 20). The best fit is $
    C_0 ~=~ - 0.016$ for both theories, and in fact regardless of the mass $m$. 
Meanwhile the best fit mass is $m=1.0$ for the SHO potential in Fig.~\ref{fig:Sk_SHO}, and  
$m=0.72$ for the PT example shown in Fig.~\ref{fig:Sk_PT}. Clearly the imaginary-time TEBD method has reproduced the desired QFT vacuum entirely as expected, modulo these two parameters. 

What, then, is the meaning of these two parameters, and why are they not equal to the ideal case, which would be $m=\omega$ and $C_0= 0$? The answer is to do with our discretization of the local Hilbert space, \ie the discrete and finite domain of $q_n$ quadratures in our qumodes. We can demonstrate this in several ways. 

First let us consider the mass $m$. It is arguably more consistent to use an effective mass-squared derived from the double derivative of a {\it fit} to the discretized potential ${\mathscr{V}}(q_n)$, rather than one derived from the double derivative of the potential itself. After all, this is more indicative of the mass-squared that is ``seen'' by the QFT in our simulation. For the SHO potential this gives a value of $m\approx 1$, while for the PT potential this gives a mass $m\approx 0.77$, both very close to the values that we had to insert into the dispersion relation in order to fit the observed spectrum $\omega(k_\alpha)$. This is very encouraging, and indeed this effective mass continues to track the curvature in the fit to the discretized potential rather than that of the ideal potential when we consider other models and when we vary the parameters.   

The parameter $C_0$ has a less immediately obvious  interpretation. From Eq.~\eqref{eq:Cnmagain} we see that it corresponds to a $\delta$-term in
$C_{nm}$  of the form  $C_{nm}\supset C_0 \delta_{nm}$. Such a term could conceivably be due to the local potential in an interacting theory, to the lattice spacing (\ie finite $a$) or again be due to the finite local Hilbert space. (Note that it could not arise due to lack of convergence as that would be $k$ dependent.) 

 In order to determine which it is, note that we have already seen that $C_0$ is independent of the particular model, \ie it is the same in the interacting PT theory as in the free-field theory with the SHO potential of Eq.~\eqref{eq:shopot}.
Moreover, when we increase the size of the Hilbert space $C_0$ begins to asymptote to zero. For example we find $C_0\approx 0.008$ when the number of quadrature points is increased from 20 to 30 (although it is computationally too demanding to go to a much more finely discretized quadrature). Therefore we can reasonably conclude that $C_0$ is also an effect of the truncated and discretized Hilbert space, and that both $m$ and $C_0$ will approach the idealised QFT values in the limit of infinite dimensionality of the qumode quadrature.  

\section{False vacuum decay in (1+1)-dimensional QFT}

\subsection{The objective: tunnelling solutions}

\label{subsec:bubble_analytic}

\begin{figure}[t!]
\centering
\includegraphics[keepaspectratio, width=0.45\textwidth]{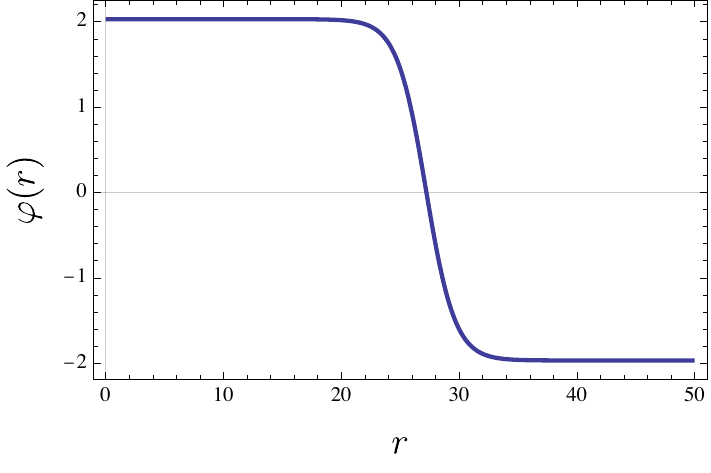}
\caption{Bubble profile for $\lambda =0.5$ and $\varepsilon =0.02$ for which Eq.~\eqref{eq:rcrit} gives $r_c=23$.}
\label{fig:bubble_profile}
\end{figure}

We now turn to the tunnelling problem that we will focus on for this study. 
For completeness let us first recap the analytic tunnelling results of Refs.~\cite{PhysRevD.15.2929,PhysRevD.16.1762,1980PhRvD..21.3305C}. In order to find the instanton tunnelling solution analytically we are required to solve the system in the 
Euclidean potential, that is with imaginary time $\tau = i t$. The solution is as we have mentioned radially symmetric in Euclidean space-time, that is it has $O(2)$ symmetry, and is hence a function of $r=\sqrt{\tau^2 + x^2} $.  
Thus, in (1+1)-dimensions the Euclidean equations of motion become
\begin{equation} \label{eq:EucEOM} \partial^2_r \varphi+r^{-1} \partial_r \varphi ~=~  \partial _\phi  {\mathscr V} ~.
\end{equation}  
This now resembles the classical equation of motion of a ball rolling in the {\it inverted} potential $-\mathscr V$, with $r$ playing the role of time and with the  $r^{-1} \partial_r \varphi$ term providing a friction that decays with time. Thus, the solution with the correct boundary conditions can be found by starting the ball close to the true vacuum at $\varphi \approx \varphi_0$ at `time' $r=0$ (corresponding to the centre of the instanton) and letting it come to a halt at $r\to\infty$ precisely at the false vacuum at $\varphi = -\varphi_0$. This requires a fine-tuning of the starting value of $\varphi$, numerically solving the one dimensional differential equation in Eq.~\eqref{eq:EucEOM}, and employing the overshoot-undershoot method  described in Ref.~\cite{PhysRevD.15.2929} to satisfy the boundary conditions. The result for the quartic potential in Fig.~\ref{fig:pot_poly} with $\lambda =0.5$, $\varphi_0=2$ and $\varepsilon =0.02$  is the bubble profile shown in Fig.~\ref{fig:bubble_profile}.
Although it is not possible to find a closed form for the  radially symmetric solution, it can be very accurately modelled by the following function:
\begin{equation}
    \varphi(r) ~\approx ~ -\varphi_0\, \tanh \beta (r-r_0)
\end{equation}
where in this case  
$r_0 = 21.8$ and the inverse width of the bubble wall is $\beta = 0.4$. The behaviour after the bubble formation is found by going back to Minkowski space time: 
\begin{equation}
    \varphi(x,t) ~\approx ~ -\,\varphi_0\, \tanh \beta (\sqrt{x^2-t^2}-r_0)~.
\end{equation}
Thus we see that once formed the bubble expands, with the bubble walls following the hyperbol\ae\/  
$x^2-t^2 = r^2_0
$
and asymptoting to the lightcones at $x=\pm t$. Outside the wall the field drops exponentially fast to the false vacuum, $\varphi = -\varphi_0$. Inside the bubble wall the field is in the true vacuum state with tiny oscillations: indeed here we can approximate 
\begin{align}
\varphi(x,t>|x|)
~\approx -\varphi_0\, (1+2 e^{-2\beta r_0}\cos{2\beta |t^2-x^2|}     ) ~.
\end{align}

What is the typical  bubble size, $r_0$? To find it when the bubble wall is thin compared to $r_0$ (the so-called thin-wall approximation) we can compare the surface tension of the bubble to the vacuum energy contained within its volume. The former can be estimated by integrating the energy across the wall,
\begin{align} 
\label{eq:sigma}
\sigma &~=~ \int dr \frac{\dot \varphi^2}{2} +  V  \notag \\
  &= \int_{-\varphi_0}^{\varphi_{\rm esc}} d\varphi \sqrt{2 \scV  }~=~ S_1~,
\end{align} 
where $S_1$ is precisely the Euclidean action of Eq.~\eqref{eq:S1}. 
A critical bubble has a radius $r_0\approx r_c$ for which the vacuum energy inside the bubble is just enough to off-set the energy of the domain walls. Using Eq.~\eqref{eq:Vp0} this implies $S_1 = 2 \varphi_0 \varepsilon r_c $ hence 
\begin{equation}
\label{eq:rcrit}
    r_c ~= ~ \frac{S_1}{2\varphi_0 \varepsilon}~.  
\end{equation}
Once formed with $r_0\approx r_c$, a bubble expands along the aforementioned hyperbol\ae\/  and if it is not impeded by friction it approaches the speed of light, driven by the  energy gained from the increasing internal volume of true vacuum. 

It is straightforward to see that, in fact, any supercritical bubble that is stationary at $t=0$  follows a hyperbola which is determined by its initial radius. Indeed, if we assume that the bubble profile is symmetric $O(2)$, then it is of the form $\varphi(x,t) = f(r-r_0)$, where $r_0$ is simply the bubble radius at $t=0$.

As mentioned in the introduction the rate of bubble nucleation per unit space-time volume is exponentially sensitive to the Euclidean action associated with the tunnelling solution. Specifically, in the thin-wall approximation  it follows the Arrhenius-like behaviour $\Gamma ~=~ \kappa \, e^{-S_1}$, where $\kappa $ is a prefactor that accounts for quantum fluctuations around the instanton solution (which is notoriously difficult to evaluate). This exponential suppression means that even small increases in $ S_1 $ can lead to a dramatic decrease in the nucleation rate. Thus the value of $S_1$ effectively encodes how ``difficult'' it is for the system to tunnel from the false vacuum to the true vacuum, with lower $S_1$ corresponding to exponentially higher nucleation probabilities and faster phase transitions.

This then is the paradigm for false vacuum decay that was laid down in Refs.~\cite{PhysRevD.15.2929,PhysRevD.16.1762}. Before considering the equivalent evolution on our qumode lattice, we should emphasize that this `classical $O(2)$ bounce' picture provides a classical core for a process that has many subtleties.  

\begin{figure*}[t!]
\centering
\includegraphics[keepaspectratio, width=0.8\textwidth]{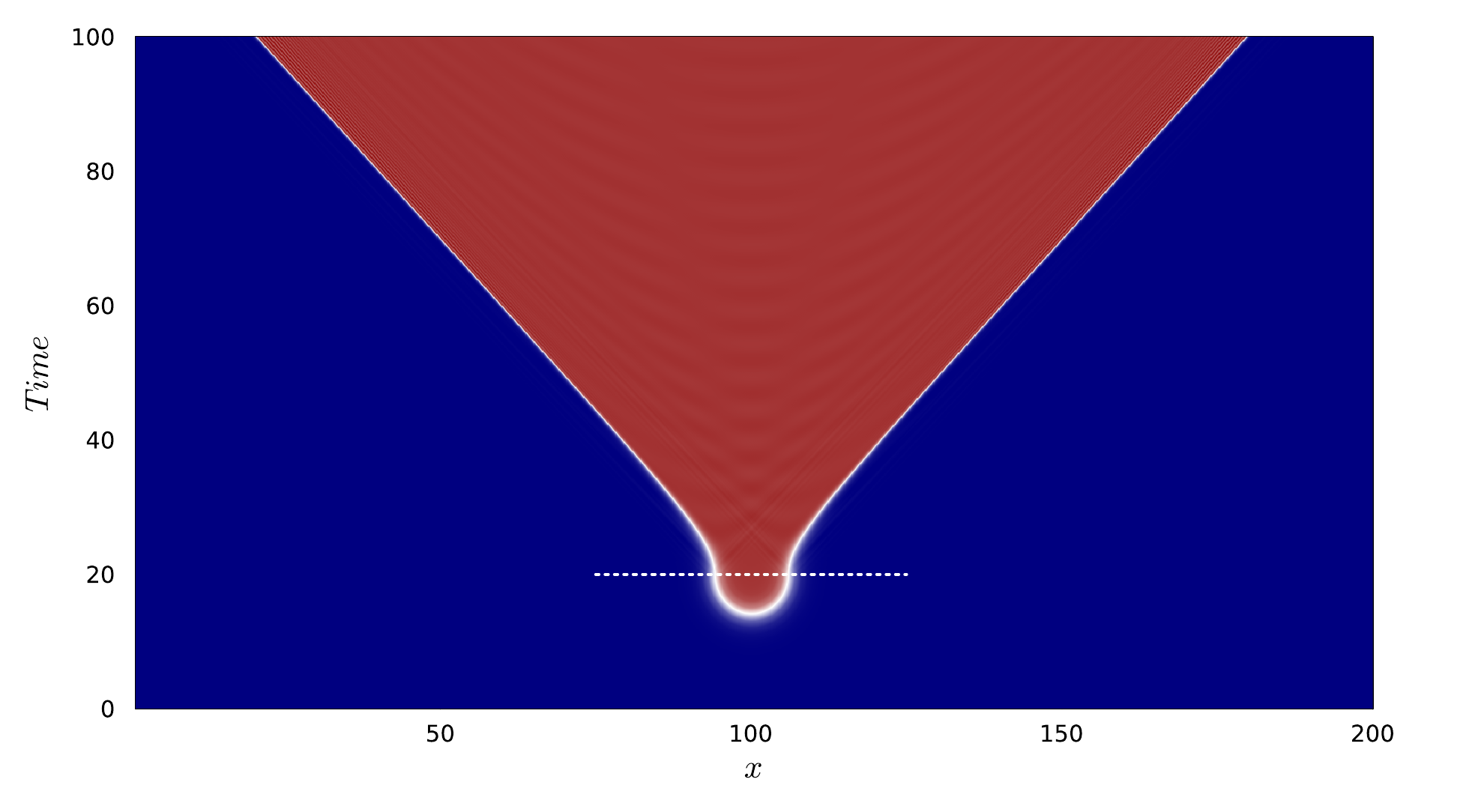}
\caption{An ideal instanton tunnelling from the metastable QFT vacuum. The system is initialised in the metastable QFT vacuum with the  $\tanh q^\ell$ potential of Eq.~\eqref{eq:tanhx2_pot} with $\lambda =\mu=1$, $\ell =2 $ and parameters  $\fv=-2$, $\tv=0.7$, $\varepsilon=0.3$. The dashed white line marks the time where the bubble appears. Below this line time is imaginary: \ie we analytically continue to Euclidean space-time which caps off the cone with a hemisphere.} 
\label{fig:instanton_ideal}
\end{figure*}

To expand on this point (and also for later comparison), let us generate the entire nucleation and growth process for a single bubble (formed at time $t_0=20$), by  numerically solving the PDEs. In Fig.~\ref{fig:instanton_ideal} we show the numerical solution to the PDE for the  $\tanh q^\ell$ potential of Eq.~\eqref{eq:tanhx2_pot} with $\lambda =\mu=1$, $\ell =2 $ and parameters  $\fv=-2$, $\tv=0.7$, $\varepsilon=0.3$. 

The solution is found by evolving from the bubble wall profile $\varphi(t_0,x)$, determined by the aforementioned shooting method. This along with $\dot\varphi (t_0,x)=0$ is our Cauchy data. We then evolve  backwards for $t<t_0$ in Euclidean time, and forwards in Lorentzian time. Thus the hyperbolic evolution after the bubble has nucleated is effectively the analytic continuation of the Euclidean $O(2)$ bounce. 

However, despite this nice global picture of the whole process, it is important to realise that the smooth Euclidean cap is {\it not} what one expects to see in pure Lorentzian time (\ie in real time). Neither does the bubble simply `pop into existence' all at once, but rather it first `appears' where the negative eigenmode of the $O(2)$ bounce has support, which is at the walls. This real-time dynamics is the quantum mechanical aspect of the tunnelling process that is most hard to capture. Thus while the $O(2)$ bounce prescription gives a useful classical core for understanding the tunnelling process it tells us little about these more subtle aspects. 

Indeed there are many other subtle issues surrounding the dynamics of these processes which have been discussed subsequently, and which are not fully understood. For example, whether or not the bubble ``runs away'' to light speed or instead reaches a terminal velocity depends on the friction generated by its interactions with the medium it is travelling through, and is a complicated and model-dependent question (see \cite{Bodecker_2009,Bodeker_2017,Hoche_2021,Gouttenoire:2021kjv} and for a recent review see \cite{Wang:2022txy}). This of course is one of the motivations for studying the process numerically.

\subsection{Results: tunnelling in (1+1)-dimensional scalar theory}\label{sec:results}

To validate the model, we simulate the real-time dynamics of the system in three distinct scenarios. First, we consider ``pre-formed'' bubbles, initialising the system with a true-vacuum bubble with an initial radius $r_0$. Second, we study phase transitions in the full system by examining bubble nucleation arising from quantum fluctuations. 

\subsubsection{Isolated bubbles: expansion and collapse}

\begin{figure*}[t!]
\centering
\begin{subfigure}[b]{0.49\linewidth}
    \centering
    \includegraphics[width=\linewidth, keepaspectratio]{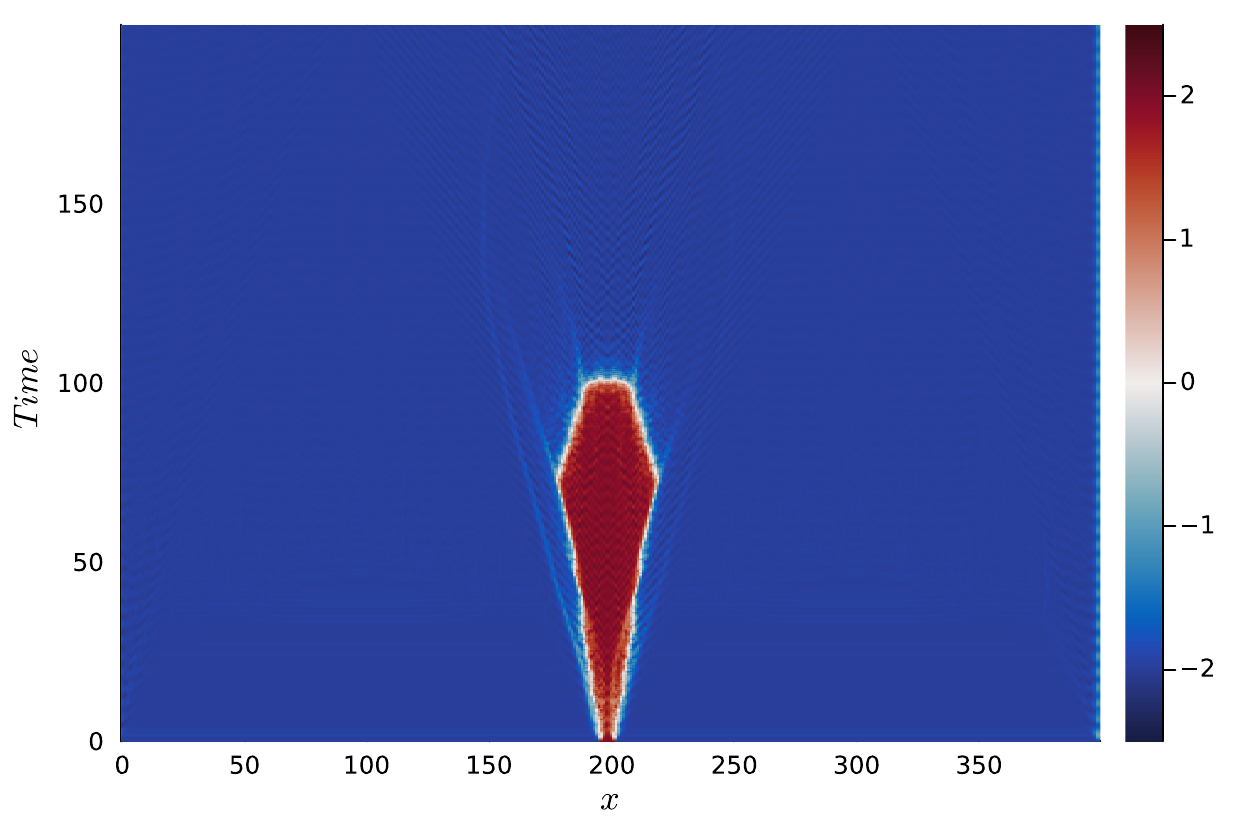}
    \caption{Subcritical bubble}
    \label{fig:seed_sub1}
\end{subfigure}
\hfill
\begin{subfigure}[b]{0.49\linewidth}
    \centering
    \includegraphics[width=\linewidth, keepaspectratio]{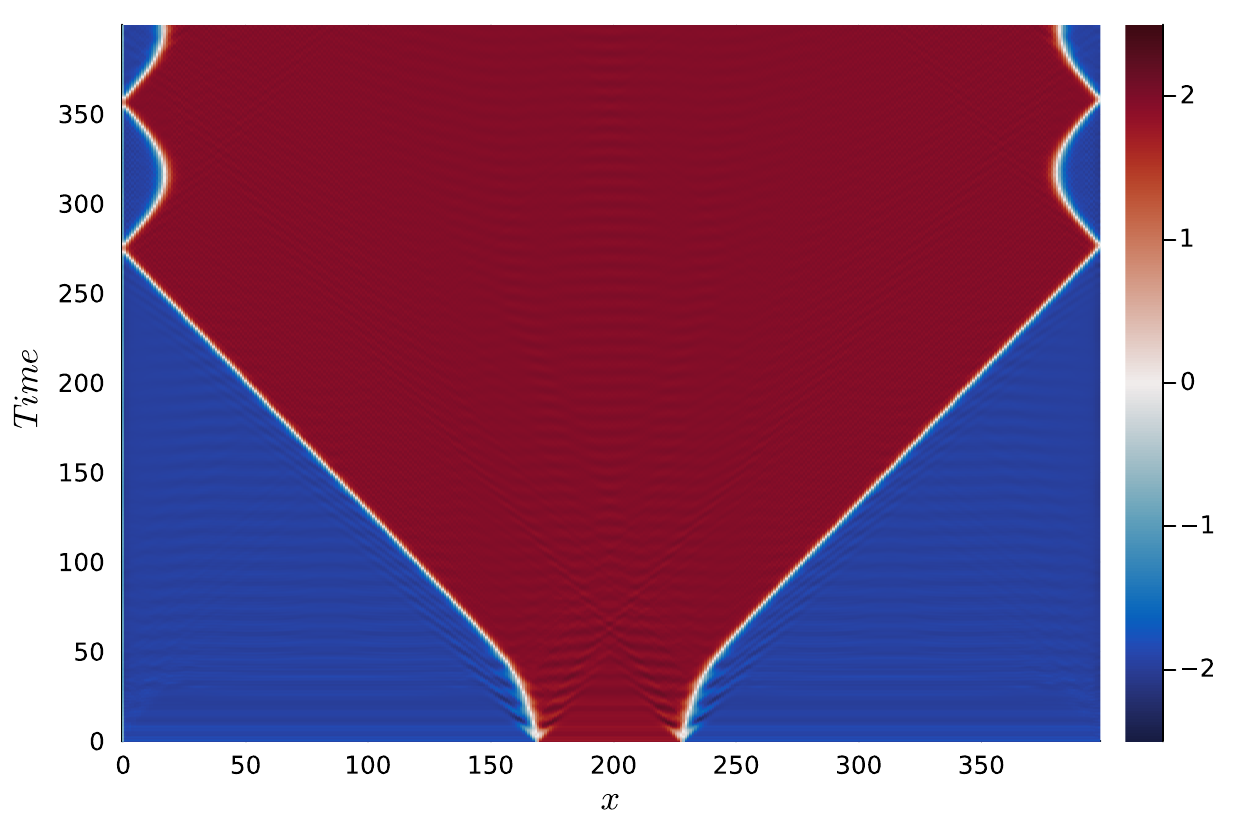}
    \caption{Supercritical bubble}
    \label{fig:seed_sub2}
\end{subfigure}
\caption{In \ref{fig:seed_sub1} we show a collapsing subcritical bubble in the P\"oschl-Teller potential of Eq.~\eqref{eq:pt_pot} with $\tv=-\fv=2$, $\alpha = 1$, $\gamma=3/2$ and $\varepsilon =0.25$. Fig.~\ref{fig:seed_sub2} shows an expanding bubble in the potential of Eq.~\eqref{eq:tanhx2_pot} with $\tv=-\fv=2$, $\lambda = 1$, and $\varepsilon =0.1$, which shows the bubble walls following a hyperbolic trajectory that reaches a terminal speed of 0.72. Note that we do not impose periodic boundary conditions, and the bubble walls ``bounce''  off the sides of the lattice elastically.}
\label{fig:expanding_tanh}
\end{figure*}

To begin, we examine the real-time dynamics of pre-formed bubbles with an initial radius, $r_0$. 
Here our initial state is taken to be sitting in the metastable ``product of qumodes'' vacuum, that is the quantum mechanical vacuum determined with no gradient-squared kinetic terms. As we have discussed, this is an approximation that is useful if long wavelength correlations are not important, which is likely to be the case when the bubble is already formed, in which case the evolution is expected to be more local and classical in nature.

Fig.~\ref{fig:expanding_tanh} presents the results from the qumode tensor network for bubbles of different initial radii. The first scenario, shown in Fig.~\ref{fig:seed_sub1}, considers the collapse of a subcritical bubble in the P\"oschl-Teller potential from Eq.~\eqref{eq:pt_pot} with $\tv=-\fv=2$, $\alpha = 1$, $\gamma=3/2$ and $\varepsilon =0.25$. In this example, we observe that the bubble fails to reach the critical radius. There are two interesting aspects of the collapse. The first phenomenon that can be seen has to do with the way the collapse occurs. We can clearly see the emission of several wavepackets from the walls. Averaged over time these emissions generate friction on the walls. Here, however, we see in addition that such emission (together with momentum conservation) is a necessary component of bubble collapse. This is because, in one spatial dimension, a single bubble wall gains no local energetic advantage if it moves to collapse the bubble, whereas in higher dimensions there is a classical energy balance between surface tension and vacuum energy, which the surface tension wins if the bubble has a subcritical radius. On a local patch of wall we see this as a nett force accelerating the patch towards the centre of the bubble if the bubble is subcritical. This classical effect is absent in one dimension because the energetic cost of surface tension is independent of the bubble radius. 
The second interesting aspect of Fig.~\ref{fig:seed_sub1} is related. It is the fact that the final moments of collapse, in which the whole bubble ``flops'' back to the false vacuum, would correspond to super-luminal wall velocity. This in turn implies that the entanglement on the qumode lattice must spread over distances of at least the radius of the collapsing bubble at that instant in time, in order for the collapse to happen. 

Figure~\ref{fig:seed_sub2} shows the expansion of a bubble in the potential from Eq.~\eqref{eq:tanhx2_pot} with $\tv=-\fv=2$, $\lambda = 1$ and $\varepsilon=0.1$. In contrast with Fig.~\ref{fig:seed_sub1}, this bubble is supercritical and we observe the phenomena outlined in Sec.~\ref{subsec:bubble_analytic}. As the bubble's radius is greater than the critical bubble radius, the walls expand. The bubble expansion is entirely driven by internal pressure caused by the energy gain from increasing the volume of the true vacuum. This acceleration is relatively local, and long-range entanglement plays very little role in the behaviour. As a result, the walls accelerate along hyperbolic trajectories which asymptote to a constant terminal velocity. However the terminal velocity is subluminal ($v_{\rm term} \approx 0.72$), due to the friction mentioned above. 

Note that we do not impose periodic boundary conditions and this results in the bubble wall colliding elastically with the sides of the lattice. We can in principle use this phenomenon to measure the mass of the bubble walls. We observe that after the bounce the walls are projected a distance of $d_{\rm bounce} \approx 20$ into the bulk of the lattice before being repelled back by the pressure, $\varepsilon$. (This relatively short distance also tells us that friction has indeed resulted in a terminal subluminal speed.) We may make a rough estimate for the mass of the wall by neglecting the effect of friction after the bounce and taking a non-relativistic approximation. Equating the initial kinetic energy with the maximum potential energy due to the pressure gives 
\begin{equation}
    m_{\rm wall} ~\approx ~ \frac{2 \varepsilon d_{\rm bounce} }{v_{\rm term}^2} ~\approx ~ 7.7~.
\end{equation}
Performing the integral in Eq.~\eqref{eq:sigma} numerically we find that the actual wall mass should be  $m_{\rm wall}~=~ S_1 ~\approx ~2.74$ for this set of parameters, so this estimate is in the right ball-park. It could potentially be improved by taking into account friction. 

\subsubsection{Phase transitions}

\begin{figure*}[t!]
\centering
\begin{subfigure}[b]{0.49\linewidth}
    \centering
    \includegraphics[width=\linewidth, keepaspectratio]{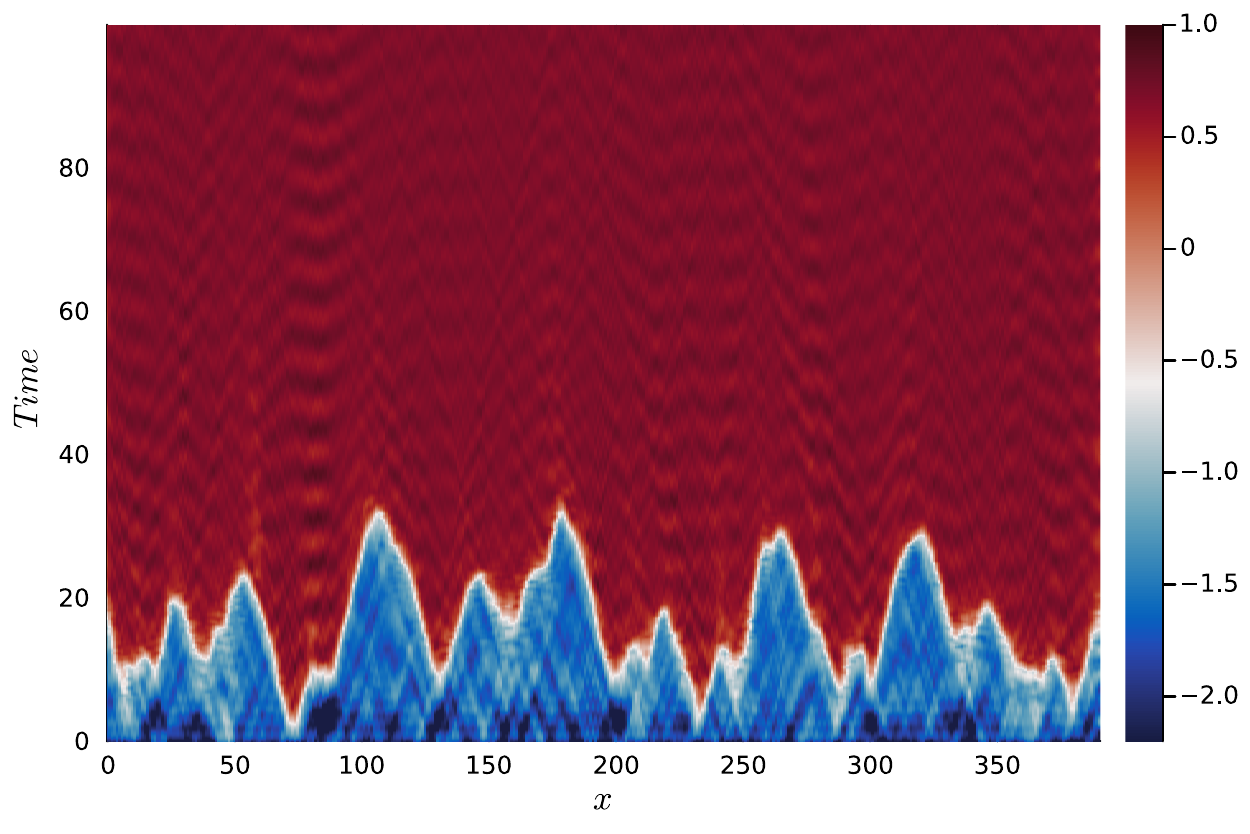}
    \caption{$\fv=-2$, $\tv=0.7$, $\varepsilon=0.5$}
    \label{fig:sub1}
\end{subfigure}
\hfill
\begin{subfigure}[b]{0.49\linewidth}
    \centering
    \includegraphics[width=\linewidth, keepaspectratio]{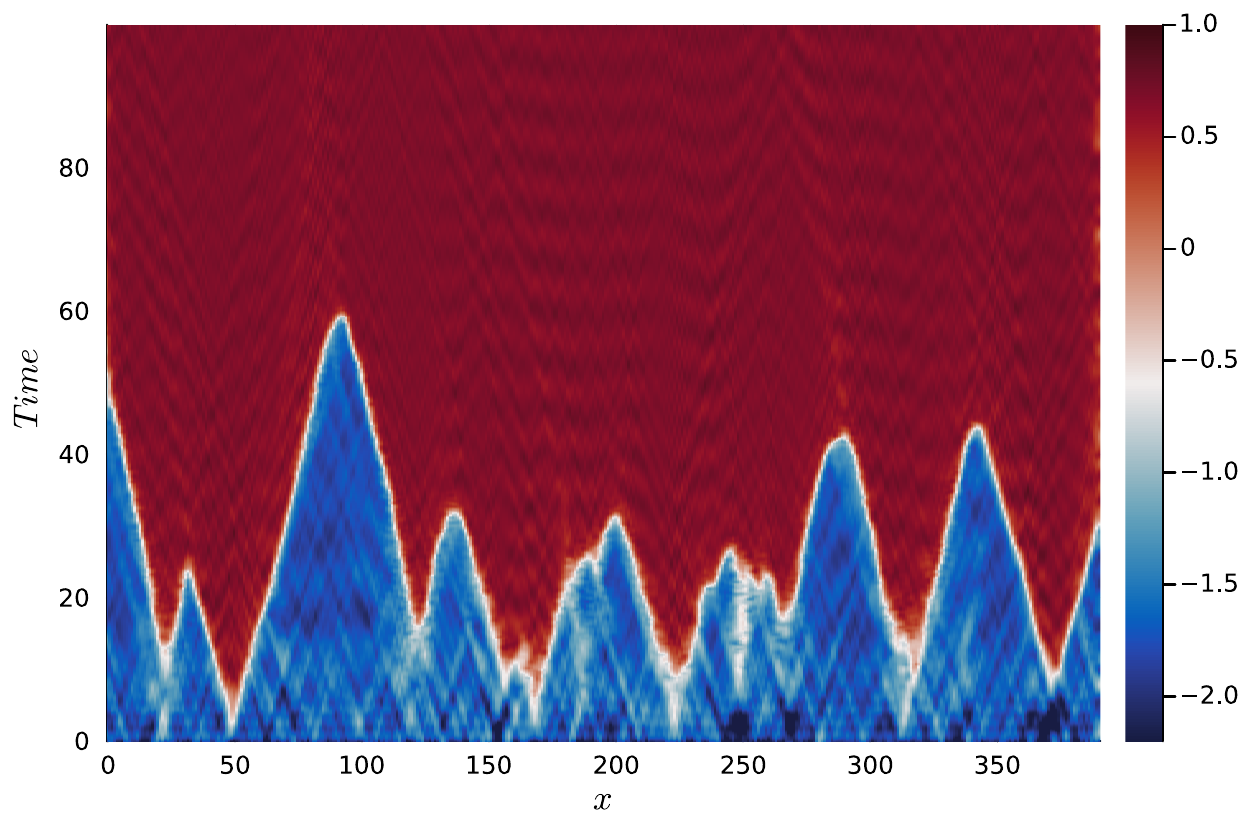}
    \caption{$\fv=-2$, $\tv=0.7$, $\varepsilon=0.4$}
    \label{fig:sub2}
\end{subfigure}
\caption{Phase transitions on the qumode lattice for two different parameter settings showing the measured expectation value $\vev{\varphi}$.}
\label{fig:lattice_transition}
\end{figure*}

Now let us turn to phase transitions in the system as a whole. In Fig.~\ref{fig:lattice_transition} we show two examples of phase transitions that were produced with the lattice initialised in the random sampling approximation to the QFT false vacuum that was discussed in Sec.~\ref{subsec:sampling}. 
The potential used for these examples is the $\tanh q^\ell$ function of Eq.~\eqref{eq:tanhx2_pot} with $\ell =2 $. The true minimum for these examples was chosen to be at $\tv=0.5$ in order to allow for relatively rapid transitions. The figures show the production of bubbles which coalesce to complete the phase transition. The first example with $\varepsilon = 0.5$ is very rapid because the instanton is very small (with $r_c\approx 2$). The second example has $\varepsilon = 0.4$ and the instanton size has grown enough ($r_c\approx 5$) to noticeably slow the bubble production, as would be anticipated from the analysis in Sec.~\ref{subsec:bubble_analytic}. One observation that supports the bubble formation being the result of genuine instanton transitions is that  we find that the maximum bond dimension has to exceed a certain threshold in order for this behaviour to appear (if the dimension is set to be too small then there is little tunnelling). Moreover, this threshold appears to grow with the instanton size. In future work, an in-depth study of these relations will be carried out, with a particular focus on confirming the exponential dependence of the decay rate $\Gamma$ on the action. 

\subsection{Seeding an instanton in the full QFT}

Finally, we present a tunnelling process in the genuine QFT metastable vacuum initialised by imaginary-time TEBD as discussed in Sec.~\ref{subsec:iTEBD}. As a test we wish to seed the production of an instanton using a small perturbation. To do this we first prepare the initial state as per Sec.~\ref{subsec:iTEBD} in the metastable QFT vacuum around $\fv$. 
We then apply a small seed to stimulate the 
nucleation of a bubble within a reasonable time. For direct comparison we will consider bubble nucleation in the $\tanh q^\ell$ potential of Eq.~\eqref{eq:tanhx2_pot} with $\lambda =\mu=1$, $\ell =2 $ and parameters  $\fv=-2$, $\tv=0.7$, $\varepsilon=0.3$. This is the same potential that was used to produce the  ideal false-vacuum decay plot in Fig.~\ref{fig:instanton_ideal}.

In order to find a suitable seed without just randomly stimulating classical (equivalently thermal) transition,  we must determine the optimal direction in field-space in which to push the field. We can do this by considering the expansion of the field around the $O(2)$ bounce solution described in Sec.~\ref{subsec:bubble_analytic}. In particular we would like to identify a negative eigenmode whose instability leads it to grow exponentially, thereby driving the decay of the false vacuum. We will refer to this as the ``negative mode''.

\begin{figure}[t!]
\centering
\includegraphics[keepaspectratio, width=0.4\textwidth]{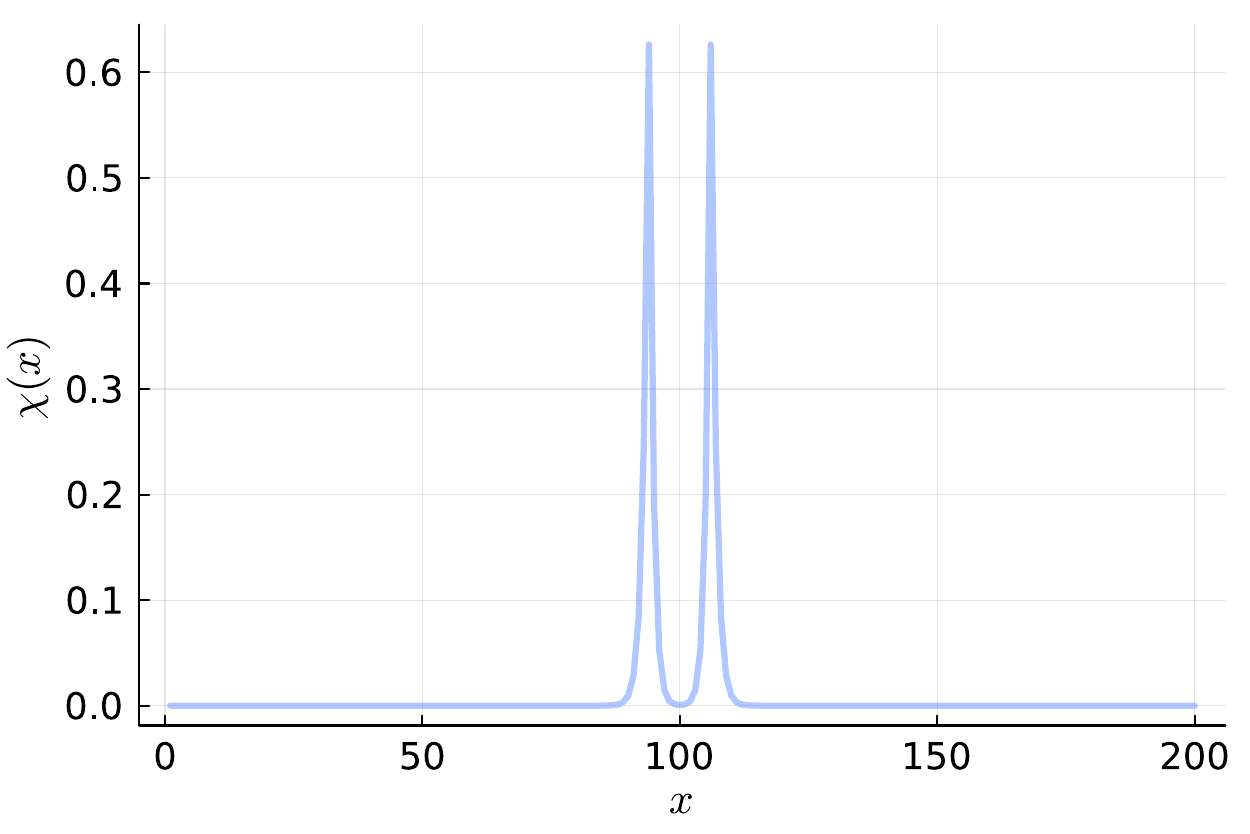}
\caption{The negative mode for false vacuum decay in the  $\tanh q^\ell$ potential of Eq.~\eqref{eq:tanhx2_pot} with $\lambda =\mu=1$, $\ell =2 $ and parameters  $\fv=-2$, $\tv=0.7$, $\varepsilon=0.3$. The mode is trapped in the ``effective potential wells'' associated with the walls of the $O(2)$ bounce solution.} 
\label{fig:growing_mode}
\end{figure}

There is a well-known procedure for this, which is to linearise the fluctuations around the bounce solution, $\varphi_B(x)$, as 
\begin{align}
\varphi(x,t) ~=~ \varphi_B (x,t) + \eta (x,t)~, 
\end{align}
such that there is an effective $x$-dependent Lagrangian for $\eta $  of the form 
\begin{align}
    {\cal L} ~=~ \frac{\eta_t ^2}{2}-\frac{{\eta}_x ^2}{2}-\frac{V''(\varphi_B(x))}{2} \eta^2 ~.
\end{align}
Separating variables,  $\eta = \chi(x)\tau(t)$, the equations of motion give  $\tau_{tt} =  \omega^2_- \tau  $, and 
\begin{align}
- \chi_{xx} ~+~ (V''(\varphi_B(x)) + \omega_-^2 )\chi~=~0~,
\label{eq:se}
\end{align}
where we choose the separation constant $\omega_-$ anticipating exponential growth of the form $\tau(t) = A e^{\omega_- t}+B e^{-\omega_- t}$. Clearly this can be consistent only if the mode in question $\chi(x)$ is a bound state in $x$ (as opposed to say an extended oscillatory mode in $x$). As Eq.~\eqref{eq:se} is simply the time-independent Sch\"odinger equation with an effective potential $V_{\rm eff}(x) = V''(\varphi_B(x))$ and energy $ \omega_-^2 $, this is only possible if there are potential wells associated with regions of negative $V_{\rm eff}$. Thus a single exponentially growing mode, namely the negative mode, will be trapped in the regions with $V''(\varphi_B)<0$, and indeed solving Eq.~\eqref{eq:se} will determine the ``ground state energy'' and hence the value of  $\omega_-$: these regions correspond to those parts of $\varphi_B(x)$ that are beneath the potential barrier, which are in turn precisely the locations of the bubble walls of the bounce solution. We show this negative mode in Fig.~\ref{fig:growing_mode}

Generally, the negative mode will grow as 
\begin{align}
    \eta(x,t) ~=~ \left(q_0 \cosh (\omega_- t )+ 
\frac{p_0}{\omega_-}\sinh (\omega_- t ) \right) ~ \chi(x) ~,
\label{eq:inst_growth}
\end{align}
with $q_0$ and $p_0$ indicating displacement and momentum at $t=0$.
Thus the desired seed consists of a small kick to the false vacuum in the direction of {\it this} mode. A suitable and straightforward implementation then is to apply a momentum kick to our vacuum MPS of the form 
\begin{align} 
\hat U _{p_0} ~=~ e^{i p_0 \chi(x_i) \hat q_i }~,
\end{align}
where $\chi(x)$ is understood to be the normalised relevant bound state solution to Eq.~\eqref{eq:se}, and where $p_0$ is a constant, which in our studies we take to be $p_0=0.1$. 

\begin{figure*}[t!]
\centering
\includegraphics[keepaspectratio, width=0.8\textwidth]{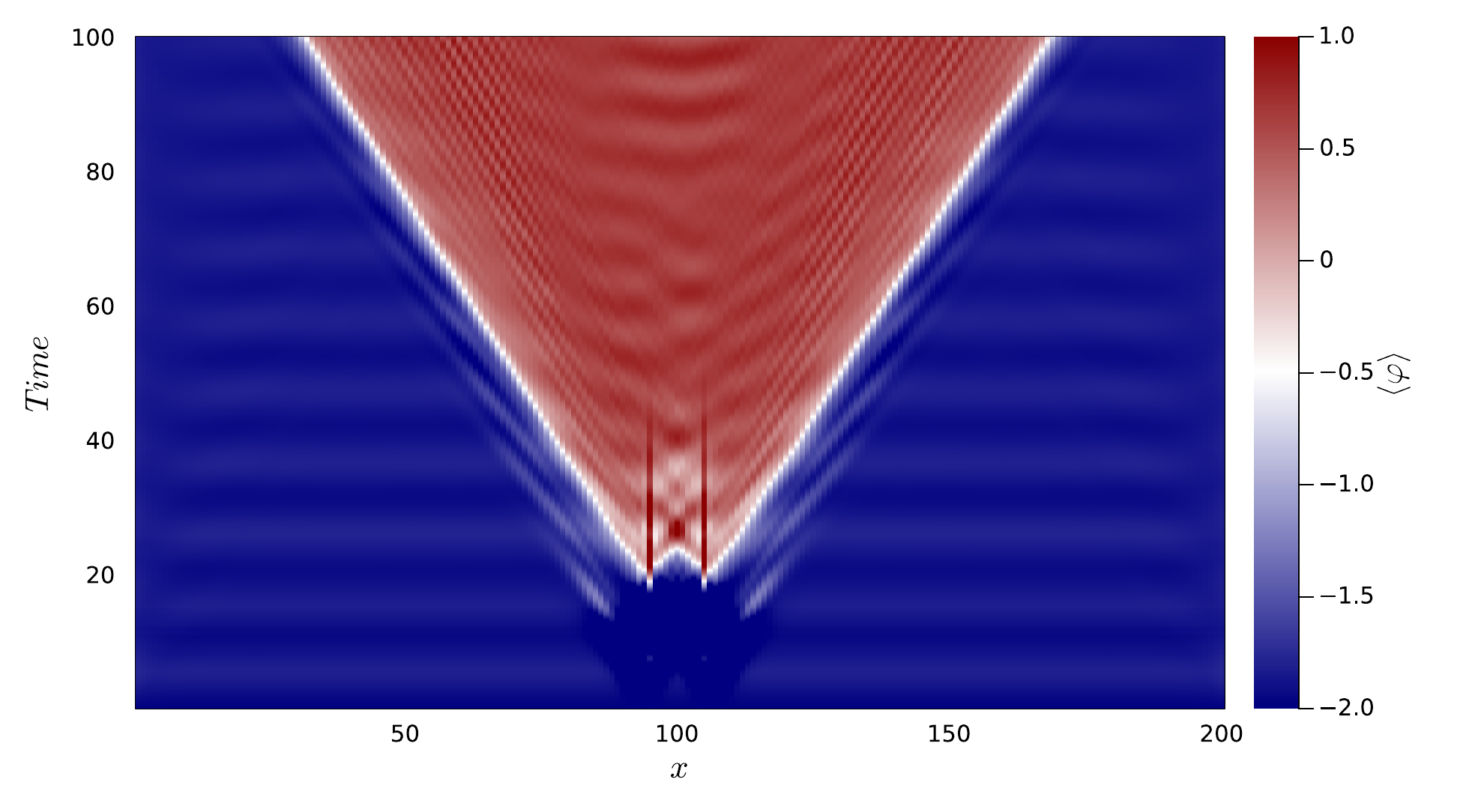}
\caption{An isolated tunnelling event from the metastable QFT vacuum. The system is initialised in the metastable QFT vacuum with the  $\tanh q^\ell$ potential of Eq.~\eqref{eq:tanhx2_pot} with $\lambda =\mu=1$, $\ell =2 $ and parameters  $\fv=-2$, $\tv=0.7$, $\varepsilon=0.3$. A small seed is applied (consisting of a small momentum boost applied at $t=0$ along the exponentially growing ``negative mode'' in Fig.~\ref{fig:growing_mode}) in order to stimulate an instantonic transition. Here $\vev{\varphi}$ is the measured expectation value.} 
\label{fig:instanton}
\end{figure*}

The subsequent evolution of the scalar field VEV is shown in Fig.~\ref{fig:instanton}, where we observe behaviour that closely follows the theoretical  prediction of Sec.~\ref{subsec:bubble_analytic} and the ideal case (with identical parameters) in Fig.~\ref{fig:instanton_ideal}.  
As expected there is an initial period during which nothing appears to happen, in which the negative mode is growing exponentially. Given the behaviour of this mode in Eq.~\eqref{eq:inst_growth}, the time for instability to set in is given by $1/\omega_-$ with only logarithmic dependence on the initial kick size $p_0$, provided that it is large enough. (As we are not actually in the background of the critical bubble, an insufficiently large $p_0$ just perturbs the system in the harmonic basin of the false vacuum, with the energy radiating away.) As anticipated the bubble appears ``walls first'', with the interior being pulled to the true vacuum by the walls. 

There are other features that we should remark on. First we note the residual negative mode which overshoots the saddle and ultimately disappears in radiation.  We can also see a residual long-lived 
``oscillon'' at the site of the decay \cite{PhysRevD.49.2978,PhysRevD.52.1920}. 

It is worth understanding the differences between the Fig.~\ref{fig:instanton_ideal} and 
Fig.~\ref{fig:instanton}
as well. The qumode lattice simulation has a somewhat less well defined hyperbolic trajectory. This is due to the discretisation of the system onto a lattice. This causes the non-ideal dispersion relation in Eq.~\eqref{eqn:omegas}, which results in a softening of short wavelength modes. (In fact we can observe the same effect when we latticise the classical PDE solution.) A second difference is the slightly enhanced oscillations in Fig.~\ref{fig:instanton}: this is  due to the seed that we introduced to stimulate the transition.

\section{Summary and Conclusions}\label{sec:conclusions}

In this work, we have presented a novel framework for simulating false vacuum decay in scalar QFT using a qumode-based Hamiltonian formulation. Our method overcomes longstanding challenges associated with simulating highly entangled, non-equilibrium tunnelling dynamics in QFT by leveraging a qumode lattice representation and implementing real-time evolution via tensor network techniques, specifically the time-evolving block decimation algorithm.

Unlike previous approaches, which typically restricted attention to spin chains or Euclidean simulations, our framework, based on Hamiltonian simulation, remains firmly within the QFT setting and directly simulates real-time tunnelling phenomena. The core innovation lies in the use of qumodes, that is bosonic modes with (quasi-)continuous Hilbert spaces, which enable the construction of scalar field theories with arbitrary potentials. This setup can, in principle, be implemented on continuous-variable quantum computing devices in its presented form. 
Simulation on current classical hardware may be performed by truncating the local Hilbert space and employing a matrix product state ansatz for the quantum state. Such truncations would not be needed on continuous-variable quantum hardware, which would enable simulations of genuine Coleman tunnelling.

This work  made two key advances. The first was an algorithm for preparing the full initial state vacuum of even strongly coupled QFTs using imaginary-time TEBD, thus enabling scalable simulations with tunable entanglement control. 
The second key advance was the identification and study of the role of the negative tachyonic mode in false vacuum decay. This goes well beyond what is possible in the traditional treatments of false vacuum decay, exposing the intrinsically quantum origin of the non-perturbative instability. 

We demonstrated the validity of our method through a sequence of other complex tests. 
Importantly, we showed that quantum bubble nucleation emerges in this framework only once sufficient entanglement is captured by the MPS. The bond dimension required increases with bubble size, suggesting a direct numerical signature of the entanglement entropy scaling expected for false vacuum decay. This provides a powerful diagnostic and highlights the necessity of entanglement-aware methods in simulating quantum phase transitions.

Compared to previous literature, our work represents the first demonstration of scalar QFT vacuum decay using a real-time evolved Hamiltonian lattice system with quasi-continuous local degrees of freedom. While continuous-variable tensor network methods, such as continuous MPS or cPEPS \cite{PhysRevX.9.021040, PhysRevD.105.045016}, have been developed for ground-state field theory, and continuous-variable quantum computing proposals exist for QFT~\cite{Abel:2024kuv, PhysRevA.109.052412}, no prior work has combined qumode-lattice discretisation with real-time tensor-network evolution to simulate false vacuum decay in QFT. Our method is thus, to the best of our knowledge, the first demonstration of qumode tensor networks applied to non-equilibrium tunnelling dynamics in scalar field theory.

It can open the door to systematic studies of vacuum decay in arbitrary scalar potentials, including non-renormalisable theories and those beyond the reach of semiclassical methods. 
Future extensions include the exploration of multi-field tunnelling landscapes, and the application of this method to higher-dimensional QFTs. Our results show the viability of qumode tensor networks as a bridge between classical simulation and future quantum computing applications in non-perturbative quantum field theory.

\bigskip

\subsection*{Acknowledgements} This work was supported by the STFC under the IPPP grant ST/T001011/1. Part of this work used the DiRAC@Durham facility managed by the ICC on behalf of the STFC DiRAC HPC Facility (www.dirac.ac.uk). The equipment was funded by BEIS capital funding via STFC capital grants ST/K00042X/1, ST/P002293/1, ST/R002371/1 and ST/S002502/1, Durham University and STFC operations grant ST/R000832/1.

\appendix

\section{Scalar field theory on a qumode lattice}\label{sec:basics}

In this Appendix, we recap the relation between scalar quantum field theories and a lattice system of coupled qumode. For this discussion a ``qumode'' can be thought of as any quantum mechanical oscillator. For the free-field theory, the qumodes in question will correspond to simple harmonic oscillators (SHOs), while in non-linear systems the qumodes will be quantum mechanical oscillators with more general potentials. 

\subsection{Free fields on the qumode lattice}

Let us begin with the Lagrangian for the system we wish to encode, namely a scalar quantum field, $\varphi(x, t)$. To establish our dictionary, it is convenient to begin with a free field, where $\varphi$ has mass $\omega$ given in natural units ($\hbar=c=1$) such that 
\begin{equation}\label{eqn:Lagrangian} 
\mathscr{L} ~=~ \frac{1}{2} \left( \partial^\mu \varphi (x, t) \partial_\mu \varphi(x, t) - \omega^2 \varphi(x, t)^2\right)~,
\end{equation}
with $\varphi$ and $\pi$ obeying the usual equal-time commutation relations. 
The conjugate momentum associated with the field is  
\begin{equation}\label{eqn:conjMomenta}
\pi(x, t) ~=~ \frac{\partial \mathscr{L}}{\partial (\partial_0 \varphi(x, t))} ~=~ \partial_0 \varphi(x, t)~, 
\end{equation}
such that the Hamiltonian density is
\begin{equation}\label{eqn:HamiltonianDens} 
\mathscr{H}(x, t) ~=~ \frac{1}{2} \left( \pi(x, t)^2 + (\nabla \varphi(x, t))^2 + \omega^2 \varphi(x, t)^2 \right)~. 
\end{equation}

We wish to simulate this system by coupling together a lattice consisting of a large number, $N$, of quantum mechanical oscillators. For simplicity, we shall henceforth focus on a one-dimensional field theory and locate the oscillators at positions $x_n = an$ where $n\in [0,N-1]$, and where $a$ is the lattice spacing. We will also enforce periodic boundary conditions throughout, such that $x_{N}= x_0$, although this boundary condition will not be consequential in our discussion. 

Each quantum-mechanical oscillator has its own quadrature operators that we denote $\hat q_n$ and $\hat p_n$ (and can initially at least be assigned its own wavefunction $\psi_n(q_n)\equiv \langle q_n|\psi_n\rangle$). It is straightforward to verify that there is a configuration of nearest-neighbour couplings of the oscillators that reproduces the free-field theory Hamiltonian, $\mathscr{H}$, in the large $N$ limit. Moreover, the field operator $\varphi(x)$ and its corresponding conjugate momentum operator $\pi(x)$ can be directly equated with the quadrature variables of the oscillators (written as a  density) at that position on the lattice, such that one can identify  
\begin{equation}\label{eqn:latticeFields}
\hat q_n(t) ~\leftrightarrow ~  \varphi(x_n, t)~, \qquad \hat p_n (t) ~\leftrightarrow ~  \pi( x_n , t)~.
\end{equation}
To make this identification, we first express the spatial derivatives that appear in the Hamiltonian density of Eq.~\eqref{eqn:HamiltonianDens}  by finite difference,
\begin{equation}\label{eqn:finiteDifference}
\partial_i \varphi(x_n, t) ~\leftrightarrow~ \frac{{\hat q}_{n+1}(t) - {\hat q}_n(t)}{a}~.
\end{equation}
We then define the total Hamiltonian by summing the contributions from the coupled individual qumodes over the entire lattice as follows:
\begin{equation}\label{eqn:TotalHamiltonian}
H ~=~ \frac{a}{2} \sum_{n=0}^{N-1} \left[   {\hat p}_n(t)^2 + \left(\frac{\hat q_{\,\overline{n+1}}(t) - \hat q_n(t)}{a}\right)^2 +  \omega^2 \hat q_n(t)^2 \right]~,
\end{equation}
with variables normalized such that 
\beq 
[ \hat q_n ,\,\hat p_m]~=~ i\delta_{nm} ~,
\eeq
and where periodic boundary conditions can be enforced if desired  using 
\begin{equation}
    \overline{n + 1} ~=~ n+1 \mod (N)~. 
\end{equation}
The $a$ prefactor in the Hamiltonian ensures that it scales extensively with the size of the lattice, ${\rm Vol}=L=aN$. We can also define the combined wavefunction and the combined set of quadrature variables, 
\beq 
\label{eq:combined}
 |\Psi \rangle ~=~ \otimes _n |\psi_n\rangle ~~;~~~ |{\mathbf q}\rangle  ~=~ \otimes _n |q_n\rangle~.
 \eeq
 In the ground state of the lattice, all the wave functions will be degenerate, and there will be no contribution to the Hamiltonian from the gradient terms:
\beq \langle {\mathbf q} | \left(\frac{\hat q_{\,\overline{n+1}}(t) - \hat q_n(t)}{a}\right)^2 | \Psi_{\rm degen} \rangle ~=~ 0~.\eeq
Evidently, $\omega$ is the natural frequency of the lattice of oscillators when the qumodes are all degenerate. The ground state energy of the system is then simply the sum of the ground-state energies of all the qumodes, $H_0 = \frac{\omega}{2}\,aN$, which is equivalently the volume multiplying the ground-state energy of a single qumode.
Perturbing this vacuum in a degenerate manner would result in a qumode lattice wavefunction  that continued to factorise into a product of independent qumodes, 
\beq \label{eq:unentangled}\langle {\mathbf q} |  \Psi_{\rm degen} \rangle ~=~ \prod_n \langle q_n |\psi_n \rangle ~.\eeq
Of course, propagating modes then correspond to perturbations of the lattice around such a degenerate ground state. However it is important to realise that a lattice of qumodes in their ground state does {\it not} correspond to the traditional QFT vacuum,  precisely due to the spatial kinetic terms, but it effectively coincides with the QFT vacuum for non-relativistic modes. We will return to the question of the QFT vacuum in detail later.

To make explicit the relation anticipated in Eq.~\eqref{eqn:latticeFields}, we must first diagonalise $H$. This is made non-trivial by the hopping terms which provide a ``circulant'' contribution. Defining 
$\nu = e^{2\pi i /N}$, the diagonalisation matrix is found to be 
\beq 
\label{eq:unitaries}
U_{\alpha m}~=~ \frac{1}{\sqrt{N} } \nu^{\alpha m}~,
\eeq
where we use Greek letters $\alpha, \beta, \ldots , \kappa $ to refer to the diagonal basis. In this diagonal basis, the Hamiltonian of Eq.~\eqref{eqn:TotalHamiltonian} becomes, 
\begin{equation}
H ~=~ \frac{a}{2} \sum^{N-1}_{\alpha =0} \left( \vert \hat{p}_\alpha \vert ^2 + \omega_\alpha ^2 \vert \hat q_\alpha  \vert ^2 \right)~,\\[1em]
\end{equation}
where \beq 
\label{eq:rotations}
\hat {p}_\alpha ~=~ \sum_n U_{\alpha n} \hat p_n~~;~~~~
\hat {q}_\alpha ~=~ \sum_n U_{\alpha n} \hat q_n~,
\eeq
and where the eigenvalues are
\begin{equation}\label{eqn:omegas}
\omega_\alpha ~=~ \sqrt{\left( \omega^2 + \frac{4}{a^2}\sin^2 \left( \frac{\pi \alpha }{N} \right) \right)}~.
\end{equation}
Thus, in the $N\to \infty $ limit we recover the usual interpretation of free-field QFT as a tower of SHO oscillators, which have momentum 
\beq 
\label{eq:NinftyK}
k_\alpha ~=~ 2\pi \frac{\alpha}{aN}~=~ 2\pi \frac{\alpha}{L}~,
\eeq
and which obey the relativistic relation $\omega_\alpha^2=\omega^2 +k_\alpha^2$.

Two remarks are in order. First note that in the diagonal basis, $\hat q_\alpha$ and $\hat p_\alpha $ are complex, with a phase given by the momentum $k_\alpha$, despite the fact that $\hat q_n$ and $\hat p_n$ are real. Consequently, each real mode in space is now represented by a {\it pair} of complex conjugate modes in momentum (Fourier) space, corresponding to left and right-moving momentum eigenstates. As their sum must be real, we can conclude that $\hat q_\alpha $ can only be considered independently observable in conjunction with momentum conservation in the QFT.

Our second remark is that it is clear from Eq.~\eqref{eqn:omegas} that the special relativistic relation $\omega _\alpha =\sqrt{ \omega^2 + k_\alpha^2 }$ is recovered only under the assumption that $\pi \alpha/N= a k_\alpha/2 \ll 1$. We can thus anticipate the possibility of non-relativistic artefacts on a finite lattice when the momenta approach unity in units of $a^{-1}$. As the lattice momentum is really $\tilde k_\alpha = (2/a) \sin ( \pi \alpha/N) $ these artifacts would resemble superluminal states that satisfy $\sin^{-1} (a \tilde k_\alpha/2) \approx a k_\alpha/2$, i.e. they would have $\alpha' = \alpha+ 2 \beta N $ where $\beta  \in {\mathbb Z}$.
Likewise, the lattice's highest physical momentum can be described as $k_\alpha \sim 1/a $. Hence the lattice-spacing $a$ defines a natural UV cut-off, and relativistic behaviour can only be modelled for modes that have $\omega a \ll 1$.

We are now able to make the direct identification promised in Eq.~\eqref{eqn:latticeFields} between $\varphi(x_n)$ and the quadrature variable at that lattice site $q_n$. To do this, recall that the ladder operators of the qumodes on the lattice are 
\begin{align}\label{eqn:ladderOps}
\hat a_n &~=~ \sqrt{\frac{\omega  }{2}} \left( \hat q_n + \frac{i}{\omega } \hat p_n\right)~, \nonumber\\
\hat a^\dagger_n &~=~ \sqrt{\frac{\omega   }{2}} \left( \hat q_n - \frac{i}{\omega } \hat p_n\right)~,
\end{align} 
normalized such that $H = \sum_n a \omega (\hat a^\dagger _n \hat a_n +\frac{1}{2})$ when all the qumodes are degenerate.
In the diagonal basis these become \begin{align}\label{eqn:momLadders}
\hat{a}_\alpha 			&~=~ \sqrt{\frac{\omega_\alpha  }{2 }}~\sum_n\left(  U_{\alpha n}\,\hat q_n + \frac{i}{\omega_\alpha} U_{\alpha n}\,\hat p_n \right)~,\nonumber\\
\hat{a}_\alpha^\dagger 	&~=~ \sqrt{\frac{\omega_\alpha }{2 }}~\sum_n\left( \hat q_n \left(U^\dagger\right)_{n\alpha} - \frac{i}{\omega_\alpha} \hat p_n \left(U^\dagger\right)_{n\alpha} \right)~.
\end{align}
The fact that the field operator at position $x_n$ is simply the quadrature variable at that point on the lattice can be seen as follows: our claim is that
\begin{align}
\varphi(x_n) &~=~ \hat q_n ~=~ \frac{1}{\sqrt{2\omega }  } (\hat a_n + \hat a_n^\dagger ) \nonumber \\
 &~=~ \sum_\alpha  \frac{1}{\sqrt{2\omega_\alpha } }\left( U_{n\alpha}^\dagger \hat{a}_\alpha  + \hat{a}^\dagger_\alpha U_{\alpha n}  \right)~,\nonumber \\
 &~=~  \sum_\alpha \frac{1}{\sqrt{2\omega_\alpha \,N}}\left( \hat{a}_\alpha e^{-i\alpha n} + \hat{a}^\dagger_l e^{i \alpha n} \right)~,\nonumber \\ 
 &~=~ \sum_\alpha \frac{1}{\sqrt{2\omega_\alpha\,N }}\left( \hat{a}_\alpha e^{-ik_\alpha x_n} + \hat{a}^\dagger_\alpha e^{ik_\alpha x_n} \right)~,
\label{eq:phi_to_q} \end{align}
where we use the $N\to \infty $ definition of the momentum in Eq.~\eqref{eq:NinftyK}. Similarly, its conjugate momentum is
 \begin{align}
\pi (x_n)   &~=~ \hat p _n ~=~  \sum_\alpha (-i) \sqrt{\frac{\omega_\alpha}{2 a N }} \left ( \hat{a}_\alpha e^{-ik_\alpha x_n} - \hat{a}^\dagger_\alpha e^{ik_\alpha x_n} \right)~.
\label{eq:pi_to_p}
\end{align}
Taking the continuum limit as $N\rightarrow \infty$, and replacing $\sum_\alpha \equiv \frac{aN}{2\pi } \int dk$ we retrieve the standard QFT definition of the field as an expansion in momentum space at a fixed time, namely 
\begin{align}
\varphi(x) &~=~ \int \frac{\textrm{d} k}{(2\pi)} \frac{1}{\sqrt{2\omega_k}}\left( \hat{a}_k e^{ikx} + \hat{a}^\dagger_k e^{-ikx} \right)~,\nonumber \\[1em]
\pi (x)   &~=~  \int \frac{\textrm{d} k}{(2\pi)} (-i) \sqrt{\frac{\omega_k}{2}} \left ( \hat a_k e^{ikx} - \hat a^\dagger_k e^{-ikx} \right)~,
\end{align}
where $\omega_k=\sqrt{\omega^2 + k^2}$, and where we have made the usual rescaling of the ladder operators by the square-root of the volume, namely $\sqrt{a} \,\hat a_\alpha \longrightarrow  \hat a_k /{\sqrt{a N}}$, in order to be consistent with  $[\hat a_k,\hat a_{k'}^\dagger ]=2\pi a \delta(k-k')$, versus   
$[\hat a_\alpha,\hat a_{\beta}^\dagger ]= \delta_{\alpha\beta}$.

\subsection{Interacting fields} 

It is straightforward now to implement an interacting theory consider the Hamiltonian density of scalar QFT in the continuum in its most general form with an arbitrary interaction, $\mathscr{V}$, such that
\begin{equation}
    \mathscr{H}(x, t) = \frac{1}{2} \left( \pi(x, t)^2 + \nabla \varphi(x, t)^2  \right) + \mathscr{V}(\varphi(x, t))~.
\end{equation}
Following the same discretisation procedure outlined above for the free field theory, on a lattice  of $N$ sites the total Hamiltonian for this system becomes,
\begin{equation}
    H ~=~ a\sum_n \left[ \frac{1}{2} \left(\hat{p}_n^2+ \left(\frac{\hat{q}_{\,\overline{n+1}} - \hat{q}_n}{a}\right)^2 \right) + \mathscr{V}_I\left(\hat{q}_n\right) \right]~.
\end{equation}
Expanding the terms within the Hamiltonian, 
we find three kinds of contribution in the total Hamiltonian, the on-site kinetic term, the on-site effective potential and the hopping terms: 
\begin{equation}\label{eqn:CVQCHam}
    H a^{-1} =  \sum_n \left[ \frac{1}{2} \hat{p}_n^2  + \mathscr{V}_{\rm eff}(\hat{q}_n) \right] - \frac{1}{a^2} \sum_n \hat{q}_{\,\overline{n+1}}\hat{q}_n~,
\end{equation}
where the on-site effective potential is $\mathscr{V}_{\rm eff}$ \begin{equation}\label{eqn:effectivePot}
    V_I(\hat q_n) ~=~ \frac{1}{a^2} \hat{q}_n^2 + \mathscr{V}(\hat{q}_n)~.
\end{equation}
In the main body of the paper we set $a=1$, but the overall factor of $a^{-1}$ can be absorbed into our definition of $\delta t$, which can  be thought of as expressing the time in lattice units.

\newpage
\bibliographystyle{inspire} 
\bibliography{refs}{}

\end{document}